\documentclass{jfm}
\usepackage{graphicx}
\usepackage{amsmath}
\usepackage{amssymb}
\usepackage{epstopdf, epsfig}
\usepackage{caption}
\usepackage{subcaption}
\usepackage{float}
\usepackage{multirow}

\usepackage{color,soul}

\usepackage[colorlinks=true,pdfborder={0 0 0},linkcolor=blue,citecolor=blue]{hyperref}

\shorttitle{Drop encapsulation and bubble bursting in capillary channels}
\shortauthor{P. Pico et al.}

\title{Drop encapsulation and bubble bursting in surfactant-laden flows in capillary channels}

\author{Paula Pico\aff{1},  L. Kahouadji\aff{1}, S. Shin\aff{2}, J. Chergui\aff{3}, D. Juric\aff{3,4} \\ \and O. K. Matar\aff{1}  \corresp{\email{o.matar@imperial.ac.uk}}}

\affiliation{\aff{1}Department of Chemical Engineering, Imperial College London, South Kensington Campus, London SW7 2AZ, United Kingdom
\aff{2}Department of Mechanical and System Design Engineering, Hongik University, Seoul 121-791, Republic of Korea
\aff{3}Universit\'e Paris Saclay, Centre National de la Recherche Scientifique (CNRS), Laboratoire Interdisciplinaire des Sciences du Num\'erique (LISN), 91400 Orsay, France
\aff{4}Department of Applied Mathematics and Theoretical Physics, University of Cambridge,  Cambridge CB3 0WA, United Kingdom}

\begin{document}

\maketitle


\begin{abstract}
We present a parametric study of the unsteady phenomena associated with the flow of elongated gas bubbles travelling through liquid-filled square capillaries under high Weber number conditions. These conditions consistently induce the formation of a re-entrant jet at the back of the bubble that commonly gives way to a deep liquid cavity. Subsequent steps include pinch-off events in the cavity to generate one or multiple encapsulated drops which may coalesce, in conjunction with the bursting of the bubble-liquid interface by either the cavity or the drops. Some of these interfacial instabilities have previously been reported experimentally \citep{Olbricht1996PORE-SCALEMEDIA} and numerically \citep{Izbassarov2016AContraction/expansion} for liquid-liquid flow in microchannels. We carry out three-dimensional direct numerical simulations based on a hybrid interface-tracking/level-set method capable of accounting for the presence and dynamic exchange of surfactants between the liquid bulk phase and the liquid-gas interface. Our results indicate that the delicate interplay amongst inertia, capillarity, viscosity, surfactant adsorption/desorption kinetics, and Marangoni stresses has a dramatic influence over the non-axisymmetric morphological structures of the encapsulated drops-elongated bubble. This strong coupling also influences the pinch-off time, penetration depth of the cavity, and number, size, and velocity of the encapsulated drops across the bubble. The observed phenomena are summarised in three main morphological regimes based on surfactant-related parameters and dimensionless groups. A discussion of the flow regime maps is also provided. 

\end{abstract}



\section{Introduction}

The problem of gas bubbles propagating in confined capillary channels has been the subject of extensive theoretical, experimental, and numerical investigation (see \cite{Etminan2021AStudies} and references therein). In addition to being a classic problem in fluid mechanics, as seen in the now-seminal works of \cite{Taylor1960DepositionSurface,Bretherton1961TheTubes,Schwartz1986OnTubes}, these flows are central to multiple applications in technology development and are found in several real-life systems. To name a few examples, gas-liquid interfaces in the micro- and capillary scale are involved in two-phase coolers \citep{Karayiannis2017FlowApplications}, ${\rm CO}_{2}$ sequestration and storage microdevices \citep{Shim2014DissolutionChannel}, obstructed pulmonary airways \citep{Heil2008TheClosure}, and volcanic conduits \citep{Carrigan1990ZoningConduits}, and have been proposed as a potential method for bacterial detachment and cleaning \citep{Khodaparast2017Bubble-DrivenMicrogeometries}. Since the early works mentioned above, extensive efforts have been undertaken to understand the phenomena dictating the bubble's behaviour along with the surrounding thin liquid film, largely through an idealisation of the operating conditions and the morphology of the bubble itself as a flat cylinder bounded by two semi-spherical menisci. It is well-known by experiments \citep{Taylor1960DepositionSurface} and theory \citep{Bretherton1961TheTubes} that the film thickness of the flat region scales as $h \sim Ca_{b}^{2/3}$ within the realms of vanishing capillary, $Ca_{b}$, and Reynolds numbers, $Re_{b}$, and as $h \sim Ca_{b}^{2/3}/(1+Ca_{b}^{2/3})$ for $Ca_{b} < 2$ \citep{Aussillous2000QuickTube,Martinez1990AxisymmetricTubes}. As inertial effects become non-negligible, \cite{Han2009MeasurementFlow} proposed a correction to the scaling exponents of $Ca_{b}$ and the introduction of the Weber number, $We_{b}$, in the scaling law expression. Particular attention has also been placed on exploring the influence of $Ca_{b}$ on the pressure drop ahead of the bubble \citep{Wong1995TheFlow}, mapping the full bubble topological profiles \citep{deRyck2002TheTube}, analysing the undulatory structures that emerge in the vicinity of the bubble rear \citep{Khodaparast2015DynamicsStudy,Magnini2017UndulationsFlows} as inertial effects begin to dominate, and predicting the final shape of the front meniscus \citep{Giavedoni1999TheTube}.

Other researchers have delved into the problem by deviating from the traditional idealised assumptions of circular and straight channels, absence of surface-active materials, fully dominating capillary forces, and rigid spherical interfaces. These studies have elucidated critical differences between the steady-state and axisymmetrical features of idealised bubbles and those that adhere to more realistic scenarios. For square or rectangular channel cross sections, the experiments of \cite{Chen2015ThinMicrochannels,Chen2016Thin-filmMethod} and numerical simulations of \cite{Magnini2020MorphologyCapillaries,Magnini2022LiquidCapillaries} have uncovered significant angular and axial non-uniformities in the liquid film thickness, whereby the effects of the rigid walls tend to flatten the interface at the channel's centre-line ($h_{min} \sim Ca_{b}$), depleting it of the liquid phase, and inducing the formation of large lobes protruding towards the channel corners in the cross-section ($h \sim {Ca_{b}^{0.445}}$). Furthermore, non-circular channels promote an exponential thinning of the liquid film along the bubble length from the nose to the rear section, as reported in \cite{Lister2006CapillaryLobes,Magnini2020MorphologyCapillaries}. The effects of other non-ideal geometries have been examined by \cite{Sauzade2018BubbleNumbers}, who carried out an experimental campaign to study the distortion of a train of bubbles in a constricting-expanding microchannel under varying $Ca_{b}$ and bubble packing conditions. This study provided evidence for the occurrence of bubble width-to-length ratio hysteresis across the sinusoidal structures of the channel, and the attenuation of these effects with decreasing $Ca_{b}$.

A wide array of investigations have focused on illustrating the effects of (in)soluble surfactants on liquid-gas confined flows, placing particular emphasis on their influence on liquid film thickness, bubble deformation, and pressure drop (see \cite{Olgac2013EffectsProblem} and references). The literature largely agrees that, under typical surfactant characteristics and negligible inertia, the liquid film displays a thickening response to surfactants of the order of 1 to $4^{2/3}$ times that predicted by \cite{Bretherton1961TheTubes}'s expression for uncontaminated interfaces. This thickening has been reported to vary with $Ca_{b}$, surface elasticity, and across the bubble length due to Marangoni-related mechanisms, which act in opposition to liquid drag forces as they push the surfactants towards the bubble back. Nonetheless, \cite{Ghadiali2003TheTube}'s numerical results in the context of pulmonary airways suggest the possibility of liquid film thinning surfactant actions under conditions of large bulk Péclet number, $Pe_{c} >> 1$, and $Ca_{b} > 10^{-1}$ due to a bifurcation and enhancement of Marangoni stresses, which drive fluid away from the thin film region and towards the bubble nose. In terms of pressure drop and bubble velocity, it has been shown that surfactants tend to increase the pressure drop across the dispersed phase \citep{Ginley1988InfluenceCapillaries, Luo2019EffectMicrochannel} and significantly reduce the bubble's mobility along its length \citep{Borhan1992EffectTubes}. \cite{Batchvarov2020EffectNumber} carried out a comprehensive computational exploration of multiple surfactant characteristics and dimensionless groups with the novelty of introducing inertial effects ($Re = 443 - 728$) to the analysis. A key finding of this work is the surfactant-induced dampening of the bubble's rear interfacial oscillatory structures, together with the evidence provided to affirm that this process is entirely Marangoni-driven rather than a consequence of localised lowered surface tension. Moreover, this investigation addressed the effect of low solubility surfactants on the bubble's dynamics, detecting a clear arrangement of the bubble's morphology into two separate regions; one fully covered in surfactants and of reduced mobility, extending from the bubble rear to a portion in the stream-wise direction that depends on $Pe_{c}$, and one region of diminished surfactant concentration that stretches to the bubble nose, characterised by mechanisms comparable to those of clean interfaces.

A noteworthy sequence of events transpires in the bubble as $Ca_{b}$ is increased to a critical value that strongly depends on its size and $Re$. As reported in a number of investigations \citep{Abubakar2022LinearTubes,Sauzade2013InitialOils,Taha2006CFDCapillaries,Giavedoni1999TheTube}, the reduced dominance of capillary forces has a distorting impact on the curvature of the front and back menisci, dragging both interfaces towards the main flow direction and causing the bubble to adopt an elongated, `bullet-like' shape. As inertial and viscous forces increase, the bubble's back undergoes a curvature inversion that allows the liquid phase to infiltrate into the bubble's domain, forming a small cavity. The fate of this cavity is highly influenced by a complex interplay of phenomena that might culminate in a continuous cavity growth in the main axial or radial direction, or in its capillary breakup to encapsulate and entrap liquid drops within the bubble. This type of encapsulation has been observed experimentally in the works of \cite{Goldsmith1963TheBubbles,Olbricht1992TheCapillary,Olbricht1996PORE-SCALEMEDIA} in liquid-liquid flows. Similar findings are reported in a limited number of numerical pursuits on multiphase confined flows, including the investigations of \cite{Tsai1994DynamicsTube} and \cite{Izbassarov2016AContraction/expansion} for contracting/expanding capillaries in Newtonian and viscoelastic fluids, respectively, \cite{Pozrikidis1992TheTube} for buoyancy-driven flows of viscous drops, \cite{Nath2017MigrationRegime} for liquid drops in creeping flow, and \cite{Andredaki2021AcceleratingRegimes} and \cite{Atasi2018InfluenceMicrochannels} for gas bubbles in the absence/presence of surfactants, respectively. The latter three studies have provided an examination of the specific conditions that induce the encapsulation phenomenon, a general elucidation of the mechanisms at play, and an overview of the encapsulation aftermath. \cite{Nath2017MigrationRegime} have remarked that, in the inertialess limit and considering fully circular channels, there exists a critical $Ca_{b}$ beyond which drop breakup and entrapment events will occur. The authors found this critical value to decrease (increase) with the initial drop-to-channel radius (dispersed-to-continuous phase viscosity). \cite{Andredaki2021AcceleratingRegimes} and \cite{Atasi2018InfluenceMicrochannels} have recently laid the groundwork for the formulation of general encapsulation and breakup regime maps for $Re<<1$, bounded by $We_{B}$ and the relationship between inertia and pressure drop across the channel in the former, and $Ca_{b}$ and the relative importance of interface surfactant adsorption/desorption in the latter case.

Despite these initial efforts, much remains unknown and unexplored about these encapsulation events. Notably, a quantitative and phenomenological description of the underlying dynamics that govern each step of the process is needed, in combination with a thorough inspection of the cavity characteristics. It is also crucial to address the multiple examples of interfacial singularities inherent to the system at hand, which have yet to be analysed from the perspective of the well-established problem of capillary breakup, from which multiple parallels naturally emerge between our system and others (e.g., contracting liquid ligaments and inkjet printing). Likewise, an analysis of the evolution of the drops beyond pinch-off is required as it is a topic much less understood than the steps that precede it \citep{Martinez-Calvo2020NaturalThreads} and could be crucial to potential applications \citep{Izbassarov2016AContraction/expansion}. Finally, and as will be seen in this paper, extending the set of surfactant parameters explored from those of \cite{Atasi2018InfluenceMicrochannels} uncovers new encapsulation and breakup outcomes, which allows us to expand and enhance the original regime maps. The main objective of this paper is therefore to carry out an extensive characterisation of the cavity formation and (post) pinch-off dynamics, taking into account the interaction between surfactants and inertia, capillarity and viscosity in a non-circular channel geometry, which correspond to non-idealities of interest, as seen in the foregoing literature review.

The rest of the paper is organised as follows: §\ref{sec:methods} introduces our numerical approach, including the governing equations, boundary conditions, solution methods, computational setup, and a validation for our system. Our results are presented and discussed in §\ref{sec:results}, where we focus first on encapsulation in clean interfaces, on the effects of surfactants next, and finalise with a formulation of encapsulation regime maps that compile all behaviours observed.


\section{Numerical methods and problem formulation}\label{sec:methods}

\subsection{Numerical modelling and non-dimensionalisation}

\begin{figure}
  \centerline{\includegraphics[scale=0.1]{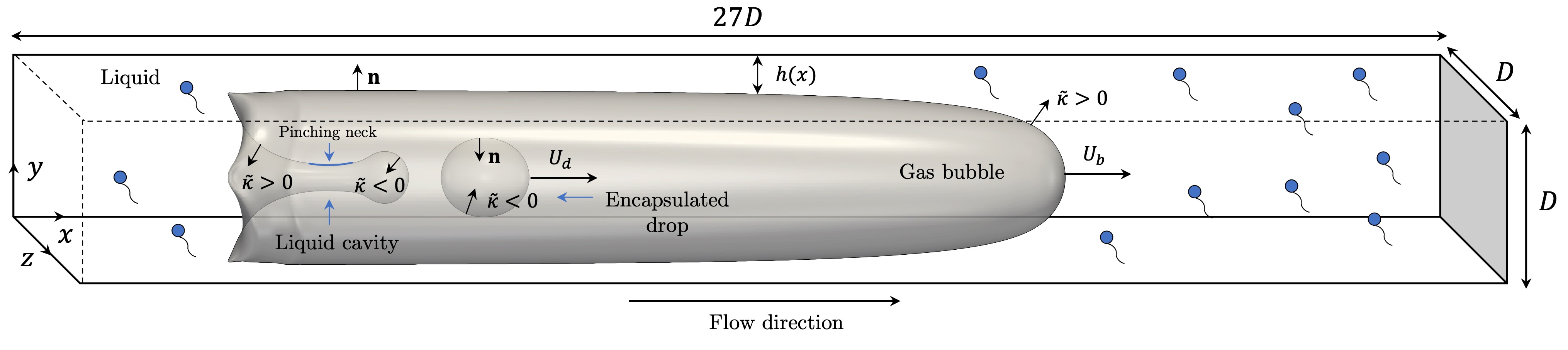}}
  \caption{(not to scale) Schematic of simulation setup, re-entrant liquid cavity at the bubble back, and encapsulated drop. The direction of outwards pointing normal vectors, liquid film thickness, $h(x)$, and curvature sign according to the adopted convention, are also highlighted. The location of the Cartesian axis represents the outset of the geometrical domain $(x=0,y=0,z=0)$}
  \label{fig_sim_conf}
\end{figure}

We consider a horizontal liquid-filled square capillary of width and height $D$ in a Cartesian three-dimensional domain, $\textbf{x} = (x,y,z)$, with an elongated gas bubble propagating through its interior (see figure \ref{fig_sim_conf}). Assuming incompressible flow, Newtonian fluids, and negligible gravitational effects, we perform direct numerical simulations (DNS) based on the two-phase Navier-Stokes equations and a hybrid front-tracking/level-set method to handle the interface and surface tension forces. This method, as formulated and implemented in \cite{Shin2002ModelingConnectivity}, is coupled with the resolution of surfactant transport and exchange between the interface and the liquid phase bulk, as described in \cite{Shin2018ASurfactant}. The relevant variables involved in the system have been rendered dimensionless (denoted by tildes) by using the scalings depicted in Eq. (\ref{eq:scaling}),

\begin{equation}\label{eq:scaling}
\begin{split}
\tilde{\textbf{x}}=\frac{\textbf{x}}{D},~
\tilde{\textbf{u}}=\frac{\textbf{u}} {U_{a}} ,~ 
\tilde{t}=\frac{t}{D/U_{a}},~
\tilde{p}=\frac{p}{\rho_{l}U^{2}_{a}},~
\tilde{\sigma} = \frac{\sigma}{\sigma_s},~\\
\tilde{\Gamma}=\frac{\Gamma}{\Gamma_\infty},~
\tilde{C}=\frac{C}{C_{\infty}},~
\tilde{C_s}=\frac{C_s}{C_{\infty}},
\end{split}
\end{equation}

\noindent where $\textbf{u}$, $t$, $p$, $\sigma$, $\Gamma$, $C$, and $C_{s}$ represent velocity, time, pressure, surface tension, interfacial surfactant concentration, bulk surfactant concentration, and bulk surfactant concentration in the vicinity of the interface, respectively. The physical parameters included in the scaling correspond to the width of the channel, $D$, the average inlet velocity of the liquid, $U_{a}$, the density of the liquid phase, $\rho_{l}$, the surface tension in a surfactant-free interface, $\sigma_{s}$, the saturation interfacial concentration, $\Gamma_{\infty}$, and the initial bulk surfactant concentration, $C_{\infty}$, in line with the scaling proposed by \cite{Batchvarov2020EffectNumber} for a similar system. In what follows, we refer to each variable by its name to refer to its dimensionless version, unless stated otherwise and add the subscript $b$ to signify that the variable is reported at the bubble nose. The governing mass and momentum equations are written in dimensionless form and according to a `one-fluid' formulation, as shown in Eq. (\ref{eq:dim_momentum})-(\ref{eq:dim_rho_mu}),

\begin{equation}\label{eq:dim_momentum}
\begin{split}
 \nabla \cdot \tilde{\textbf{u}}=0, ~~~~
 \tilde{\rho}\left(\frac{\partial \tilde{\textbf{u}}}{\partial \tilde{t}}+\tilde{\textbf{u}} \cdot\nabla \tilde{\textbf{u}}\right)  &= - \nabla \tilde{p}  + \frac{1}{Re}\nabla\cdot  \left [ \tilde{\mu} (\nabla \tilde{\textbf{u}} +\nabla \tilde{\textbf{u}}^T) \right ] \\
 &+ \frac{1}{ReCa} \int_{\tilde{A}(\tilde{t})} \left(
\tilde{\sigma} \tilde{\kappa} \textbf{n} 
 +  
 \nabla_s  \tilde{\sigma}  \right)\delta \left(\tilde{\textbf{x}} - \tilde{\textbf{x}}_{_f}  \right) d\tilde{A} , 
 \end{split}
\end{equation}

    \begin{equation} \label{eq:dim_rho_mu}
    \left.\begin{array}{c}
    \tilde{\rho}\left( \textbf{x},t\right)=\rho_g + (\rho_{l} - \rho_{g}) \mathcal{I}\left( \textbf{x},t\right),\\
    \tilde{\mu}\left( \textbf{x},t\right)=\mu_{g} + (\mu_{l} - \mu_{g}) \mathcal{I}\left( \textbf{x},t\right),\\
    \end{array}\right.
    \end{equation}
    
\noindent where $\mathcal{I}\left( \textbf{x},t\right)$ is a smoothed Heaviside function that adopts the value of zero in the gas phase and unity in the liquid phase, $\rho$ is the density and $\mu$ the viscosity of the fluids. Here, we denote the liquid and gas phases with the subscripts $l$ and $g$, respectively. The normal and tangential components of the surface tension forces are represented by the last two terms on the right-hand-side (RHS) of Eq. (\ref{eq:dim_momentum}), wherein $\tilde{\kappa}$ corresponds to twice the mean interface curvature, $\textbf{n}$ to a unit normal to the interface, $\delta \left(\tilde{\textbf{x}} - \tilde{\textbf{x}}_{_f}  \right)$ to a Dirac delta function that is zero everywhere except for the interface (located at $\tilde{\textbf{x}} = \tilde{\textbf{x}_{_f}}$), $\tilde{A}(\tilde{t})$ to the time-dependent interface area, and $\nabla_s$ to the surface gradient operator \citep{Shin2009ATechniques,Shin2017A}. We adopt the convention that positive $\textbf{n}$ vectors point outwards from the interface towards the liquid phase. Accordingly, positive values of $\tilde{\kappa}$ describe a convex interface, as exemplified in figure \ref{fig_sim_conf}. The mass conservation equations of surfactant species at the interface and bulk are given by Eq. (\ref{eq:dim_Gamma})-(\ref{eq:dim_c}) and the source term representing surfactant exchange between the interface and the bulk region immediately adjacent is given by Eq. (\ref{eq:bulksource_nd}):

 \begin{equation} \label{eq:dim_Gamma}
 \frac{\partial \tilde{\Gamma}}{\partial \tilde{t}}+\nabla_s \cdot (\tilde{\Gamma}\tilde{\textbf{u}}_t)=\frac{1}{Pe_s} \nabla^2_s \tilde{\Gamma}+ Bi \left ( k  \tilde{C_s} (1-\tilde{\Gamma})- \tilde{\Gamma}  \right ),
 \end{equation}
\begin{equation} \label{eq:dim_c}
\frac{\partial \tilde{C}} {\partial \tilde{t}}+\tilde{\textbf{u}}\cdot \nabla \tilde{C}= \frac{1}{Pe_c} \nabla\cdot(\nabla \tilde{C}),
\end{equation}

\begin{equation}
\label{eq:bulksource_nd}
\textbf{n}\cdot\nabla \tilde{C} |_{interface}=-Pe_c Da Bi \left ( k  \tilde{C_s} (1-\tilde{\Gamma})- \tilde{\Gamma}  \right ),
\end{equation}

\noindent where $\tilde{\textbf{u}}_{t} = (\tilde{\textbf{u}}_{s} \cdot \textbf{t})\textbf{t}$ is the projection of the interface velocity vector, $\tilde{\textbf{u}}_{s}$, on the interface unit tangent, $\textbf{t}$. The dependence of surface tension on local interface surfactant concentration is represented by a non-linear equation of state derived from Langmuir adsorption isotherm, as expressed in Eq. (\ref{eq:sigma}) 

\begin{equation}
\label{eq:sigma}
\tilde{\sigma}=\max\left(\epsilon_{\sigma}, 1 + \beta_s \ln{\left(1 -\tilde{\Gamma}\right)}\right),
\end{equation}

\noindent where $\beta_{s} = \mathcal{R} T \Gamma_{\infty}/\sigma_{s}$ is the surfactant elasticity number; $\mathcal{R}$ and $T$ are the thermodynamic ideal gas constant and temperature, respectively. The non-linear equation of state described above is valid for very dilute systems in which $\Gamma << \Gamma_{\infty}$. As $\Gamma$ increases, the equation of state yields unphysical negative values of surface tension. To circumvent this, a limiting value for $\tilde{\sigma}$, $\epsilon_{\sigma} = 0.05$, has been introduced, in accordance with \cite{Muradoglu2014SimulationsFlow} and \cite{Shin2018ASurfactant}. The dimensionless groups that appear in the above equations and that characterise the system are defined as follows:

\begin{equation}\label{eq:dimless}
\begin{split}
Re=\frac{\rho_lU_{a} D}{\mu_l};~Ca=\frac{\mu_{l}U_{a}}{\sigma_s};~
Pe_c=\frac{U_{a} D}{D_c};~Pe_s=\frac{U_{a} D}{D_s};~ Bi=\frac{k_d D}{U_{a}}; \\ ~Da=\frac{\Gamma_\infty}{D C_{\infty}};
~k=\frac{k_a C_{\infty}}{k_d},
\end{split}
\end{equation}

\noindent where $Re$ and $Ca$ are the Reynolds and Capillary numbers, respectively ($We = CaRe$, and $Oh = \sqrt{Ca/Re}$). $Pe_{c}$ and $Pe_{s}$ are the bulk and interfacial Péclet numbers, which provide a measure of the importance of inertia relative to surfactant mass diffusion across the interface (modulated by a diffusivity $D_{c}$) or the bulk (modulated by a diffusivity $D_{s}$), respectively; $Bi$ represents the Biot number, describing the competition between convective, $t_{conv} = D/U_{a}$, and interface surfactant desorption, $t_{des} = k_{d}^{-1}$, time scales. The Damköhler number, $Da$, provides a dimensionless measure of the initial interface saturation, and the number $k$ reflects the interplay between the characteristic time scales of interface surfactant adsorption, $t_{ad} = (k_{a}C_{\infty})^{-1}$, and desorption.

For an exhaustive description of the discretisation schemes and numerical framework employed for solving the above mentioned governing equations, we refer the reader to \cite{Shin2018ASurfactant,Shin2017A}, and to \cite{Shin2009ATechniques} for a comprehensive account of the hybrid front-tracking/level-set method, also called Level Contour Reconstruction Method (LCRM). Here, we present a short summary of the most relevant aspects. The LCRM considers the combination of a fixed structured Eulerian grid for the resolution of field equations and a moving and deforming Lagrangian grid for the interface, discretised via an unstructured triangular mesh. These interface elements are consequently advected through the integration of the Lagrangian equation $\text{d} \textbf{x}_{f}/\text{d}t = \textbf{v}$, where $\textbf{v}$ corresponds to the interface velocity, which is interpolated from the Eulerian grid. This integration is carried out with a second-order Runge-Kutta method. The spatial derivatives of the fields in the Eulerian grid are discretised through a standard cell-centred scheme for all terms, with the exception of the non-linear convective terms, for which a second-order
essentially non-oscillatory (ENO) procedure is used \citep{Shu1989EfficientII,Sussman1998AnFlows}. For the viscous term in the momentum equation, a second-order centred difference scheme is employed. With regard to the temporal terms, a second-order Gear method \citep{Wang2006GearSystems} is adopted with an implicit time integration for the viscous terms. The time step is set to be adaptive, according to the following criterion: $\Delta t = \min\{\Delta t_{\rm{cap}}, \Delta t_{\rm{vis}}, \Delta t_{\rm{CFL}}, \Delta t_{\rm{int}}\}$, where the time steps relate to viscosity, capillarity, Courant-Friedrichs-Lewy condition, and interface, respectively. The time-step related to the interface was found to be the limiting time-step in all of our simulations ($\Delta t_{\rm{int}} \sim $ \textit{O}($10^{-7}$)s).

\subsection{Simulation setup and validation}\label{subsec:sim_setup}

As depicted in figure \ref{fig_sim_conf}, our simulation domain consists of a three-dimensional rectangular box of dimensions $27D \times D \times D$, corresponding to the $x$ (channel length), $y$ (channel height), and $z$ (channel width)-directions, respectively. The value of the channel length was selected to ensure the attainment of a steady-state bubble propagation for $We < 10$, following the results of \cite{Batchvarov2020EffectNumber} and \cite{Magnini2020MorphologyCapillaries}. Under conditions of $We > 10$, highly unsteady encapsulation/bursting phenomena are detected in the system. The channel length chosen allowed for the observation of all relevant phenomena, including the formation of the liquid cavity, drop encapsulations and coalescence events, and bubble bursting, for all conditions tested. The elongated bubble was initialised in quiescent conditions close to the channel inlet and at the centre-line of the $y$ and $z$-directions as a horizontal cylinder of length $3.12D$ with two hemispherical caps of diameter $0.94D$ on each end. A fully-developed liquid velocity profile was imposed at the channel inlet with a Neumann condition for pressure ($\partial p/\partial \textbf{n} = 0$). All walls were treated as no-slip boundaries, and Neumann conditions were imposed for all variables at the outlet. The bubble-liquid interface was initially covered in surfactant with uniform concentration $\Gamma_{0}$, determined from an initial equilibrium state between surfactant adsorption and desorption (last term on the RHS of Eq. (\ref{eq:dim_Gamma}) equal to zero), which, assuming $C_{s} = C_{\infty}$ initially, reads as:

\begin{equation}\label{eq:eq_gamma}
\begin{split}
\Gamma_{0} = \frac{k}{k+1}\Gamma_{\infty}.
\end{split}
\end{equation}

\noindent This set-up closely resembles the approaches followed in previous numerical investigations for similar systems \citep{Atasi2018InfluenceMicrochannels,Nath2017MigrationRegime,Luo2019EffectMicrochannel}.

We explore the effect of the dimensionless numbers and flow parameters that characterise the system (see Eq. (\ref{eq:dimless})) considering the following ranges: $Re = 100 - 443$, $Ca = 0.0089 - 0.0693$ ($We = 3.94 - 30.70$, $Oh = 4.48\times10^{-3} - 1.25\times10^{-2}$), $Pe_{c,s} = 100$, $Bi = 0.10 - 1$, $Da = 0.01 - 1$, $k = 0.10 - 10.00$, and $\beta_{s} = 0.25 - 1$. For all cases simulated, the viscosity and density ratio between the phases is kept constant and in line with representative values for water and air: $\rho_{l}/\rho_{g}\sim$ \textit{O}($10^{3}$) and $\mu_{l}/\mu_{g}\sim$ \textit{O}($10^{2}$). We define a `base' case from which our parametric sweep was carried out with the following conditions: $Re = 443$, $Ca = 0.0693$, $Pe_{c,s} = 100$, $Bi = 0.10$, $Da = 1$, $k = 0.10$, and $\beta_{s} = 0.50$. Unless stated otherwise, all results correspond to simulations under these conditions. The selection of the testing ranges was based on common values encountered in bubbly flow systems in the presence of surfactants, as reported in \cite{Atasi2018InfluenceMicrochannels}, as well as values of $We$ that would allow a significant disruption of the bubble back to generate the aforementioned re-entrant cavity and consequent drop encapsulation \citep{Giavedoni1999TheTube,Andredaki2021AcceleratingRegimes}. For typical ionic and non-ionic surfactants in water, such as sodium dodecyl sulphate (SDS), N-dodecyl-N,N-dimethylammonio-3-propane sulfonate, and Triton X-100 (TX100), the value of $\Gamma_{\infty}$ ranges between {\textit{O}}($10^{-6}-10^{-5}$) mol/m$^{2}$ and $D_{s}$ between {\textit{O}}($10^{-12}-10^{-8}$) m$^{2}/$s, which results in $Pe_{s}=\textit{O}(10^{3}-10^{6})$ \citep{Kalli2022ComparisonMicrochannels,Constante-Amores2021DynamicsInterface}. As noted previously in \cite{Constante-Amores2021RoleCoalescence} and \cite{Batchvarov2020EffectNumber}, saturation of the system's dynamics is reached above $Pe_{s} \sim 100$. $Pe_{c}$ was set to be equal to $Pe_{s}$, following the study of \cite{Agrawal1988SurfaceTheory}.

Based on the work of \cite{Kamat2018RoleCascades} and \cite{Constante-Amores2021RoleCoalescence} and our range of characterising $Oh$, we define an inertial-capillary or Rayleigh time scale as $t_{in-cap} = \sqrt{\rho_{l}D^{3}/\sigma_{s}}$ for $Oh << 1$ and a Marangoni time scale as $t_{\rm{m}} = \sqrt{(\mu_{l}D)/\Delta \sigma}$, where $\Delta \sigma = \sigma_{s} - 0.05\sigma_{s}$ is a measurement of the surface tension gradients. The ranges for the characteristic time scales of the system are estimated to be: $t_{ad}\sim$ \textit{O}($10^{-4}-10^{0}$)s, $t_{des}\sim$ \textit{O}($10^{-3}-10^{-1}$)s, $t_{in-cap}\sim$ \textit{O}($10^{-3}-10^{-2}$)s, $t_{conv}\sim$ \textit{O}($10^{-3}-10^{-2}$)s, and $t_{m}\sim$ \textit{O}($10^{-4}$)s (for surfactant-laden cases). These time scales ensure that sorptive/desorptive, inertial, capillary, and Marangoni phenomena will play an important role in the dynamics of the system. 


Following numerous studies making use of the numerical methods described above for inertial and capillary phenomena (see \cite{Constante-Amores2021DirectInteractions,Batchvarov2020EffectNumber,Constante-Amores2020DynamicsNumber}), we employ a fully-structured and uniform Cartesian grid divided into $54 \times 4 \times 4$ sub-domains, further divided into $64 \times 32 \times 32$ cells per sub-domain, rendering a global resolution equal to $3456 \times 128 \times 128$ cells. On account of ensuring this resolution produces mesh-independent results, we have performed a mesh independence study in surfactant-free and surfactant-laden conditions, detailed in Appendix \ref{ap:mesh} along with other resolution considerations. The mesh size selected also allowed to capture accurately the thin liquid film surrounding the bubble, which, according to the works of \cite{Gupta2009OnMicrochannels} and \cite{Magnini2020MorphologyCapillaries}, requires a minimum of 5-10 computational cells covering its domain for a correct development of the liquid velocity profile. In our simulations, we have ensured the placement of at least 5-6 cells in the liquid film section for the cases that exhibit encapsulation ($We \sim 30$), for which $h(x)/D \sim 5^{-2}$ (see figure 11 of \cite{Magnini2020MorphologyCapillaries}). This condition is further achieved in the surfactant-laden simulations, which develop a thicker liquid film. We make the note, however, that the size of the entire domain, as well as the comprehensive sweep over dimensionless numbers presented, make it prohibitively expensive in terms of computation to achieve the refinement levels reported in works such as \cite{Castrejon-Pita2015PlethoraFilaments,Li2016CapillaryTransitions} to fully capture all regime transitions leading to the interfacial singularities of pinch-off.

\begin{figure}
  \centerline{\includegraphics[scale=0.062]{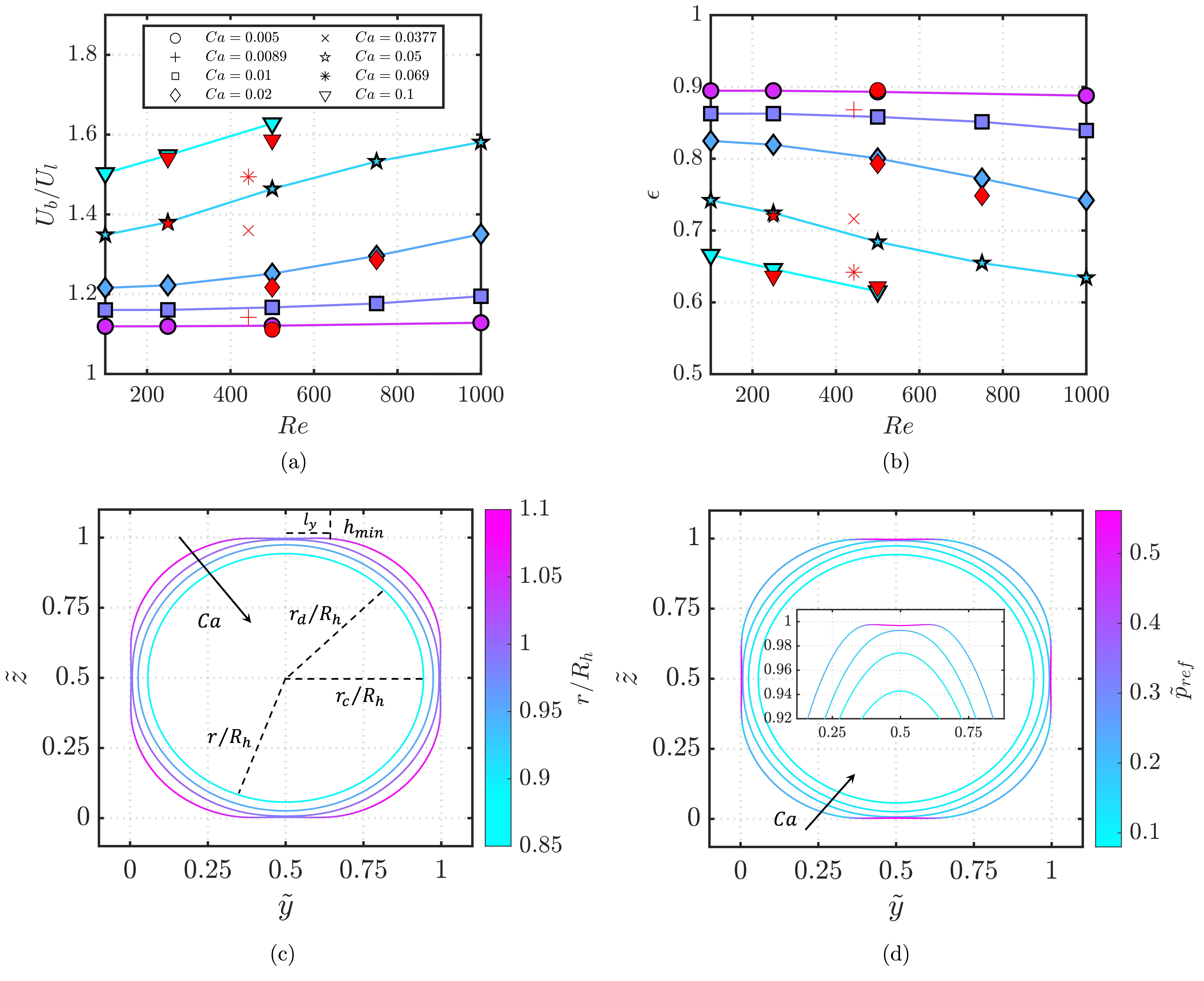}}
  \caption{Validation for square capillary channels and $We > 1$. (a) Bubble nose velocity relative to average liquid phase velocity. (b) Gas phase area fraction. Each marker corresponds to a different $Ca$, where the colour red represents our numerical results and the remaining colours \cite{Magnini2020MorphologyCapillaries}'s results. (c) Distance from channel centre to any point of interface normalised by the channel's hydraulic radius, $R_{h}$. (d) Normalised relative pressure, $\tilde{p}_{ref} = (p - p_{b})/\rho_{l} U^{2}_{a}$. Results in (b)-(d) are measured at a cross-sectional plane normal to the stream-wise direction located at a distance of $5.5D$ behind the bubble nose, as reported in \cite{Magnini2020MorphologyCapillaries}. From outermost to innermost series: $Ca = 0.0089$, $Ca = 0.02$, $Ca = 0.0377$, and $Ca = 0.1$ for $Re = 443 - 500$ in (c) and (d)}
  \label{fig_validation}
\end{figure}

The numerical framework and simulation setup presented herein have previously been carefully validated in the context of elongated bubbles in circular capillary channels. For this validation, we refer to the work of \cite{Batchvarov2020EffectNumber}, in which the code was benchmarked against the well-known correlation of \cite{Han2009MeasurementFlow} for steady-state liquid film thickness in conditions of non-negligible inertia. For $Ca \sim$ \textit{O}($10^{-2}$) and $Re \sim$ \textit{O}($10^{2}$), deviations of up to $10\%$ were reported, which are within the uncertainty of the correlation. A supplementary validation for square capillary channels is conducted using the numerical data and conditions described in \cite{Magnini2020MorphologyCapillaries} for surfactant-free scenarios under non-negligible inertia. Satisfactory quantitative agreement between our results and those of \cite{Magnini2020MorphologyCapillaries} is achieved in terms of bubble-to-liquid velocity ratio, $U_{b}/U_{l}$, and gas area fraction, $\epsilon$ (see figure \ref{fig_validation}(a)-(b)), with maximum deviations of 2.7\% and 3.1\%, respectively. We underline that the cases that do not have a direct counterpart from the above mentioned reference (i.e., $Ca = 0.0089$, $Ca = 0.0377$, and $Ca = 0.068$ for $Re = 443$) adequately follow the qualitative trend of increasing $U_{b}/U_{l}$ and decreasing $\epsilon$ with $Ca$ and $Re$.

Figures \ref{fig_validation}(c)-(d) exemplify the well-known bubble morphological regimes found in non-circular channels, as well as our model's correct capturing of the non-axisymmetric bubble shapes that arise below a threshold $Ca$. \citep{Kolb1991CoatingSection,Hazel2002TheCross-section,Chen2015ThinMicrochannels}. In accordance with \cite{Ferrari2018NumericalMicrochannels}, the corner flow effects exhibited in $Ca \lessapprox 0.01$ promote the development of a saddle-shaped interface nearby the channel centre lines (see outermost series in figures \ref{fig_validation}(c)-(d)), which brings about the formation of four lobes projecting towards the channel corners. As portrayed in figure \ref{fig_validation}(c), these lobes are located furthest away from the channel centre-point (highest $r/R_{h}$) and will continue to grow as the saddle deepens ($r_{d} > r_{c}$) with decreasing $Ca$. The presence of the saddle shifts the minimum liquid film thickness, $h_{min}$, from the centre-line of the channel to a position, $l_{y}$, that varies as a function of $Ca$. We compute $l_{y}/R_{h} = 0.2$ for the case highlighted in the figure, which matches remarkably well with the value reported by \cite{Magnini2020MorphologyCapillaries} for $We << 1$ ($l_{y}/R_{h} \approx 0.195$, see figure 7(b) of the reference). The high pressure zones at the saddles showcased in figure \ref{fig_validation}(d) are evidence of significant capillary flow drainage away from the thin-liquid film region, as reported in \cite{Magnini2020MorphologyCapillaries}.

The other morphological regimes represented in figures \ref{fig_validation}(c)-(d) correspond to an additional non-axisymmetric cross-section regime for $0.01 \lessapprox Ca \lessapprox 0.05$, characterised by $r_{d} > r_{c}$ and the absence of the saddle ($l_{y} = 0$, see $Ca = 0.02$ in the figure), and a fully symmetric bubble topology ($r_{d} = r_{c}$, see $Ca = 0.1$ in the figure) for the final regime. We draw attention to the almost perfect alignment of these regimes and the ranges of $Ca$ in which they occur between our numerical predictions and those of the literature, notwithstanding that the regimes documented in the literature were observed under conditions of vanishing inertia.


\section{Results and discussion}\label{sec:results}

Following the description of the numerical methods, simulation setup, and validation, the results of our parametric study are presented in three main parts. The first (§\ref{subseq:results_clean}) describes and analyses the underlying mechanisms responsible for the changes in curvature of the interface at the bubble tail and nose observed for increasing $Ca$ in surfactant-free cases. These changes in curvature result in a re-entrant cavity and subsequent drop encapsulation, which are illustrated in detail. §\ref{subseq:results_surfactants} expands on these concepts by incorporating into our analysis of encapsulation the effect of several different surfactant parameters related to adsorption/desorption kinetics. Finally, §\ref{subseq:results_deformation} explores the multiple non-axisymmetric structures that develop at the highly-deformed bubble tail after encapsulation for cases where adsorption phenomena are dominant.

\subsection{Bubble dynamics in surfactant-free cases: tail undulations, curvature disruptions, and encapsulations} \label{subseq:results_clean}

\begin{figure}
  \centerline{\includegraphics[scale=0.11]{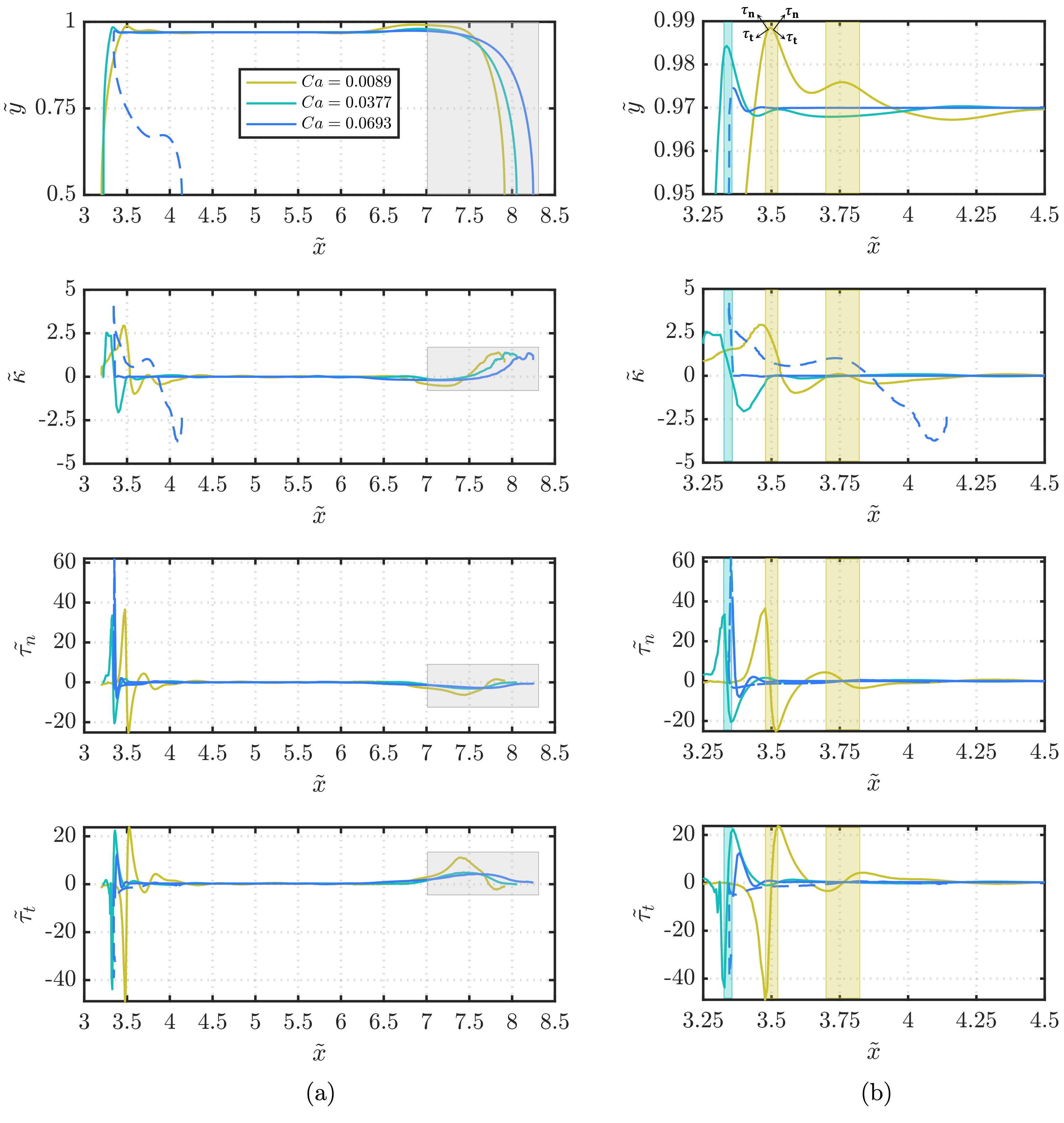}}
  \caption{Effect of $Ca$ on bubble characteristics for surfactant-free cases at $\tilde{t} = 0.89$. (a) Full bubble length in the axial direction. (b) Zoom over the trailing undulatory structures at low $Ca$ ($3.25 < \tilde{x} < 4.5$). From top to bottom the plots correspond to bubble-liquid interface shape, normalised mean curvature, $\tilde{k}$, normalised normal component of viscous stress on the interface, $\tilde{\tau}_n$, and normalised tangential component of viscous stress on the interface, $\tilde{\tau}_t$, respectively. Dotted line for $Ca = 0.0693$ highlights the re-entrant cavity. All other parameters remain unchanged from those specified in §\ref{subsec:sim_setup} for the base case}
  \label{fig_effect_ca}
\end{figure}

We start our discussion by showcasing the bubble topology at a two-dimensional projection in the ($x$,$y$) plane at the channel centre-line ($\tilde{z} = 0.5$), as well as the normalised mean curvature, $\tilde{k}$, and the normal, $\tilde{\tau}_n$, and tangential, $\tilde{\tau}_t$, components of the viscous stress along the interface (see figures \ref{fig_effect_ca}(a)-(b)). Here, these quantities are defined as $\tilde{k} = k/k_{c}$, $\tilde{\tau}_{n} = \tau_{n}/(D/U_{a})$, and $\tilde{\tau}_{t} = \tau_{t}/(D/U_{a})$, respectively, where $\tau_{n} = (\mathsfbi{D} \cdot \textbf{n}) \cdot \textbf{n}$, $\tau_{t} = (\mathsfbi{D} \cdot \textbf{n}) \cdot \textbf{t}$, $\mathsfbi{D} = (\nabla \textbf{u} + \nabla \textbf{u}^{T})/2$ (rate of deformation tensor), and $k_{c}$ corresponds to the curvature of a sphere with a volume equal to the initial bubble volume. Under the sign conventions adopted, $\tilde{\tau}_{n} > 0$ denotes stresses exerted on the interface towards the liquid phase. Conforming to previous observations, such as those by \cite{Taha2006CFDCapillaries,Giavedoni1999TheTube,Sauzade2013InitialOils,Abubakar2022LinearTubes}, which uncover the prominent impact of $Ca$ on bubble shape, our plots reveal a loss of sphericity (typical of fully dominant capillary forces) at the bubble front and back with increasing $Ca$, in conjunction with an overall axial bubble elongation. The interface at the bubble front sharpens and expands, whilst the bubble back becomes progressively flattered with $Ca$, adopting a `bullet-like' shape ($\tilde{\kappa}_{front} > \tilde{\kappa}_{back}$). This process continues until a curvature inversion emerging from $\tilde{y} = 0.5$ is detected for $Ca = 0.0693$, which eventually leads to a drop encapsulation.

Detailing the behaviour of $\tilde{\kappa}$ for $Ca = 0.0693$ along the interface path going from the cavity vertex to the bubble nose, a zone of negative curvature is first observed at $3.8 \lessapprox \tilde{x} \lessapprox 4.1$, followed by a positive curvature meniscus at $3.3 \lessapprox \tilde{x} \lessapprox 3.8$. This particular curvature sign inversion is a manifestation of the early stages that precede pinch-off and drop encapsulation, in which the fragmenting liquid filament undergoes the formation of a discernible pinching neck with a vanishing radius. This process, and its connection to the well-established scaling laws, are examined further along in this paper. Exiting the interior of the liquid cavity towards the region adjacent to the liquid film ($3.3 \lessapprox \tilde{x} \lessapprox 7.0$), a non-uniform low curvature zone that encompasses the majority of the bubble is identified, later culminating in the high-curvature bubble nose seen for $\tilde{x} > 7.0$. Contrasting this evolution to that of the two lowest $Ca$, it can be noticed that the spatial location that marks the beginning of the bubble nose (sudden curvature rise from the thin-liquid film region, highlighted in gray in figure \ref{fig_effect_ca}(a)) shifts to lower $\tilde{x}$ for decreasing $Ca$. Notable curvature inversions at the bubble tail are also identified for the low $Ca$ cases, expressed in the undulatory structures emphasized in figure \ref{fig_effect_ca}(b). These interfacial waves, previously reported in \cite{Magnini2017UndulationsFlows,Edvinsson1996Finite-ElementFlow,Giavedoni1999TheTube}, moderately diminish in frequency and amplitude with $Ca$ at the same $Re$, as illustrated in the figure.

The dynamics of the bubble-liquid interface and its curvature gradients are readily mapped onto the $\tilde{\tau}_{n,t}$ spatial profiles of figure \ref{fig_effect_ca}. Pertaining to the trailing undulatory structures (shown in yellow and green for $Ca = 0.0089$ and $Ca = 0.0377$, respectively), it is found that these are first characterised by a sudden surge in $\tilde{\tau}_{n}$ that terminates in a local maximum, which pushes the interface towards the channel walls and thus creates the crests of the interfacial waves. These peaks are promptly followed by a rapid $\tilde{\tau}_{n}$ decrease approaching a local minimum that contributes to the formation of the wave troughs by pulling the interface back towards the bubble. Concurrently, the tangential stresses feature inverse patterns of local maxima/minima that promote the sharpening of the crests/troughs. A substantial drop in $\tilde{\tau}_{n,t}$, mirrored by $\tilde{\kappa}$, is observed by departing from the rear undulatory region to the high-pressure central section of the bubble. $\tilde{\tau}_{n,t}$ remain quasi-constant and close to zero throughout this section until the reappearance of local maxima/minima marks the onset of the bubble nose. Note the generally higher $|\tilde{\tau}_{t}|$ exerted on the bubble nose at $\tilde{y} = 0.5$ for decreasing $Ca$, partly contributing to a lower curvature (more spherical) bubble front. Similar remarks about the close relationship between curvature, capillarity, and interfacial stresses have been made by \cite{Atasi2018InfluenceMicrochannels}.

\begin{figure}
  \centerline{\includegraphics[scale=0.43]{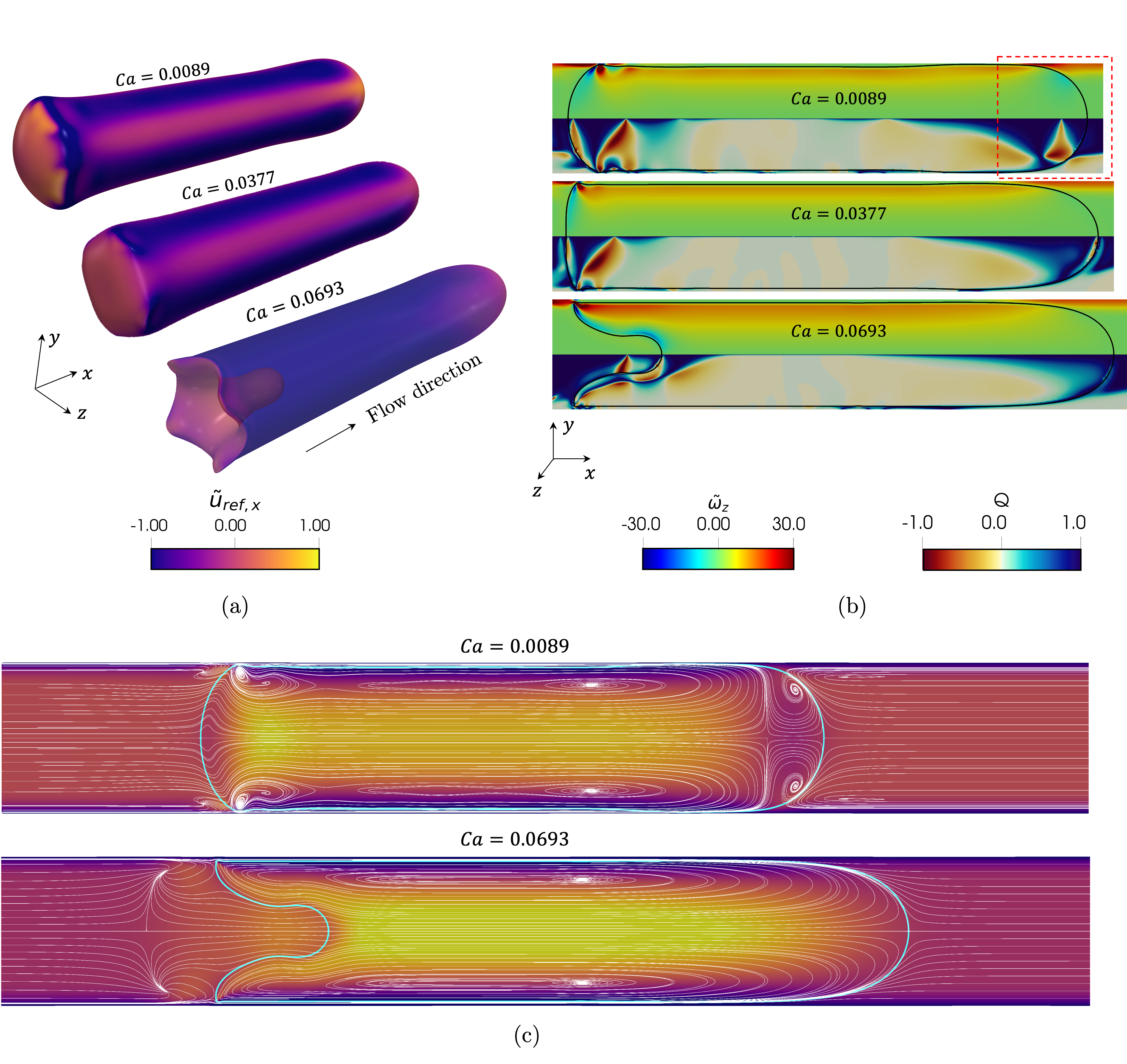}}
  \caption{Effect of $Ca$ on velocity characteristics for surfactant-free cases at $\tilde{t} = 0.89$. (a) Three-dimensional relative velocity in x-direction, $\tilde{u}_{ref,x} = (u_{x} - U_{b})/U_{b}$. $U_{b}/U_{a} = 1.0679, 1.2284, 1.3272$ for $Ca = 0.0089,0.0377,0.0693$, respectively. (b) Normalised azimuthal vorticity (top) and flow topology parameter (bottom). (c) Relative bubble velocity in x-direction and streamlines drawn in the bubble-tip reference frame. Contour plots on two-dimensional projection in x-y plane ($\tilde{z} = 0.5$) in (b) and (c). All other parameters remain unchanged from those specified in §\ref{subsec:sim_setup} for the base case}
  \label{fig_omega_Q_ca}
\end{figure}

Figure \ref{fig_omega_Q_ca}(a) shows a three-dimensional representation of the interface at $\tilde{t} = 0.89$, coloured by the $x$-component velocity relative to the bubble nose, $\tilde{u}_{ref,x} = (u_{x} - U_{b})/U_{b}$. Figure \ref{fig_omega_Q_ca}(b) depicts the azimuthal component of the normalised vorticity field, $\tilde{\omega}_{z} = \omega_{z}/(D/U_{a})$ (top half), and the flow topology parameter, $Q$ (bottom half) as contour plots over an $(x,y)$ plane for the three $Ca$ studied. Here, $Q$, which measures the influence of the bubble-liquid interface on the flow field and relates this information to the dominant flow type, is defined as follows
%
\citep{Soligo2020EffectTopology} 
\begin{equation}\label{eq:Q}
        Q = \frac{\mathsfbi{D}^{2} - \mathsfbi{O}^2}{\mathsfbi{D}^{2} + \mathsfbi{O}^2} = \begin{cases}
             -1 & \mbox{for purely rotational flow} \\
             0 & \mbox{for pure shear flow} \\
             1 & \mbox{for purely elongational flow},
        \end{cases}
    \end{equation}
%
where $\mathsfbi{D}^{2} = \mathsfbi{D}:\mathsfbi{D}$, $\mathsfbi{O}^{2} = \mathsfbi{O}:\mathsfbi{O}$, and $\mathsfbi{O} = (\nabla \textbf{u} - \nabla \textbf{u}^{T})/2$ 
corresponds to the 
rate of rotation tensor. 
Insights into the velocity distribution along the interface, and how this relates to the overall bubble motion across the channel can be gained from the three-dimensional structures displayed in figure \ref{fig_omega_Q_ca}(a). These images show that, although the bubble nose velocity increases in magnitude with $Ca$ (and $We$) due to increasingly prominent inertia, a large portion of the interface for $Ca = 0.0693$ remains notably slower than the bubble nose. In particular, a quasi-uniform spatial distribution corresponding to $u_{ref,x} < 0$ is observed in the regions adjacent to the thin liquid film. In contrast, $Ca = 0.0089$ exhibits a sequential pattern of large positive $u_{ref,x}$ at the diagonal cross-section lobes and negative $u_{ref,x}$ at the saddle-like, high-pressure regions (refer to figures \ref{fig_validation}(c)-(d)). These differences in velocity distribution are attributed to corner flow effects, which drive the cross-sectional interfacial non-uniformity at low $Ca$, leading to a very thin film ($h(x)/D \sim 3.1\times 10^{-3}$) surrounding the saddles with $u_{ref,x} < 0$ and a thicker ($h(x)/D \sim 1.7\times 10^{-1}$) liquid layer neighbouring the lobes with $u_{ref,x} > 0$. Notice that the liquid layer of the diagonal lobes in $Ca = 0.0089$ is significantly thicker than its uniform counterpart in the high $Ca = 0.0693$ ($h(x)/D \sim 5.0\times 10^{-2}$) case, explaining the negative $u_{ref,x}$ found in high $Ca$.

Upon inspection, the contour plots of figure \ref{fig_omega_Q_ca}(b) reveal a strong coupling between high-vorticity zones, such as those represented by the trailing undulations for the two lowest $Ca$, and purely rotational flow areas. Correspondingly, the low-vorticity zones seen close to the channel centre ($\tilde{y} = 0.5$) are associated with either pure shear or purely elongational flow. The back undulations portray alternating arrangements of clock- and anticlockwise vortices that decrease in $|\tilde{\omega}_{z}|$ as the amplitude of the waves diminishes. For the highest $Ca$ simulated, only one set of clock- and anticlockwise vortices are seen at the bubble back edge. Focusing on the areas near the bubble nose, a crucial difference between the three cases is identified and illustrated by the $|\tilde{\omega}_{z}|$ decrease observed for increasing $Ca$ (see red highlight in the figure). The effects of this are seen in the progressive shrinkage of purely rotational flow areas and their substitution by shear and elongational flows for high $Ca$. These larger areas of pure elongation/shear as opposed to rotation provide partial evidence to explain the axially longer bubbles that result from higher $Ca$.

To further investigate this point, we refer to the streamlines depicted in figure \ref{fig_omega_Q_ca}(c) for the two limiting $Ca$, where a few key differentiating features between the two cases are unveiled. From left to right we first recognise two stagnation points near the channel walls and almost reaching the bubble rear for $Ca = 0.0089$. These stagnation points are also observed for $Ca = 0.0693$, albeit detached from the interface and closer to the channel centre. The trailing interface undulations identified for low $Ca$ are emphasised by the vortical streamlines displayed, markedly absent in high $Ca$. Moving along the $x$-axis, we arrive at the core section of the bubble, which features a pair of central vortices covering much of the regions nearby the interface. Tying in the above analysed rotation patterns and dominating flow phenomena at the bubble nose to the arrangement of the streamlines, we point to the two vortices that develop at the bubble nose in $Ca = 0.0089$, and that have vanished in $Ca = 0.0693$. Lastly, we note that the streamlines in the liquid phase ahead of the bubble represent the occurrence of a `bypassing effect' at high $Ca$, in which the streamlines crossing the thin liquid layer travel in the main flow direction without recirculating back to the bubble, as seen at low $Ca$. The above described features of the streamlines surrounding the bubble, and how these are influenced by $Ca$, are consistent with the findings of \cite{Taha2006CFDCapillaries,Abadie2013MixingMicroreactors,Atasi2018InfluenceMicrochannels}.

\begin{figure}
  \centerline{\includegraphics[scale=0.085]{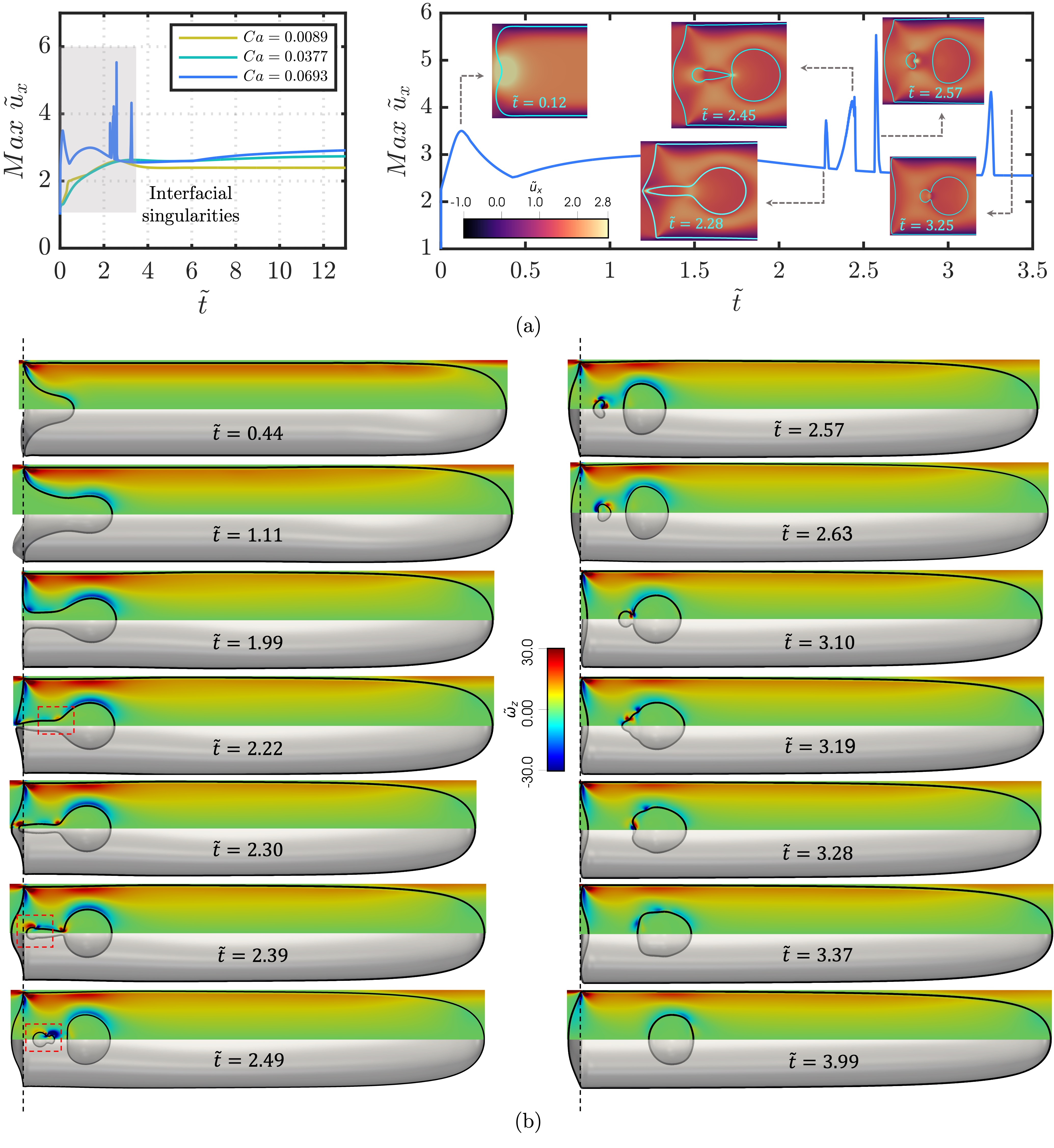}}
  \caption{Temporal evolution of interface characteristics in surfactant-free cases (a) Maximum dimensionless velocity in x-direction (left) and zoom on interfacial singularities found for $Ca = 0.0693$ (right). (b) Detailed evolution of encapsulation process for $Ca = 0.0693$. Contour plots of normalised azimuthal vorticity (top) and three-dimensional bubble shape (bottom). Refer to the Supplementary material for an animation of the encapsulation process}
  \label{fig_temporal_evolution_Ca}
\end{figure}

We now proceed to detail the encapsulation process by first illustrating the numerous flow singularities produced by high $Ca$ conditions in uncontaminated interfaces. In figure \ref{fig_temporal_evolution_Ca}(a) we plot the temporal evolution of the maximum normalised velocity in x-direction, $Max$ $\tilde{u}_{x}$. This evolution reveals five major local peaks in $Ca = 0.0693$ in the interval $0 < \tilde{t} < 3.5$, which materialise in (i) the previously analysed curvature inversion at the bubble back, (ii)-(iii) two pinch-offs that give rise to a `satellite' and a primary drop, (iv) an abrupt retraction and subsequent expansion of the former, and (v) a coalescence event between the two drops (see figure \ref{fig_temporal_evolution_Ca}(b)). A first description of the mechanisms by which these interfacial singularities come about is shown in figure \ref{fig_temporal_evolution_Ca}(c), where a cavity deepening in the positive $x$-direction takes place at $0.44 < \tilde{t} < 1.11$. The fate of this liquid cavity can be understood through the hybrid lens of the well-known problems of liquid filament breakup surrounded by a gas phase \citep{Eggers2008PhysicsJets,Castrejon-Pita2012BreakupFilaments} and liquid dripping faucets \citep{Ambravaneswaran2004Dripping-jettingFaucet}. Our liquid cavity at $\tilde{t} < 1.11$ can be thought of as a quasi-cylindrical inviscid ($Oh = 0.0125$) liquid filament with a low ratio between its length, $2L_{cavity}$, and radius, $R_{cavity}$. As in the systems studied by \cite{Wang2019AFilaments}, the radius of our cavity is not small enough for viscous drag effects from the gas phase to be significant ($R_{cavity} \sim \textit{O}(10^{-4})m> \mu_{c}\mu_{d}/\rho_{c}\sigma_{s}$) nor is our local $We$ characterising the gas phase ($We_{d} = We_{c}\rho_{d}/\rho_{c} \sim \textit{O}(10^{-2}$), $We_{c} = \rho_{c}U^{2}_{cavity}R_{cavity}/\sigma_{s}$) large enough for gas inertial effects to be influential ($We_{d} > 0.2$) \citep{vanHoeve2010BreakupJets,Lin1998DropJet}. Considering this parameter space, we are within the 'dripping' or 'jetting' regimes, as defined in \cite{vanHoeve2010BreakupJets}.

At around $\tilde{t} = 1.11$ in figure \ref{fig_temporal_evolution_Ca}(c), we observe a halt in cavity axial growth as capillary pressure builds at the bubble rear, inducing a vertical expansion towards the channel centre-line ($\tilde{y} = 0.5$), creating an essentially flat bubble back at $\tilde{t} = 1.99$ and a nascent pinching neck characterised by a high $|\tilde{\omega}_{z}|$ ring. The interface areas covered by the ring are further pulled in the counter-flow direction, prompting an additional curvature inversion at the bubble rear and the first pinch-off thereupon (see $\tilde{t} = 2.22$). This capillary singularity promotes a swift and opposite-direction recoil of the bubble back and the newly-entrapped liquid structure, thus forming the globoid-shaped bubble rear observed in $\tilde{t} = 2.39 - 2.63$ (notice the change in direction of rotation from negative to positive $\tilde{\omega}_{z}$ at the pinching neck between $\tilde{t} = 2.22$ and $\tilde{t} = 2.30$). The rapidly forming rear pinching neck witnesses the incipience of an additional capillary neck at the interior of the bubble (see highlight and high $\tilde{\omega}_{z}$ ring in $\tilde{t} > 2.22$). This concludes in a second pinch-off and a precipitous recoil into a satellite drop (see $\tilde{t} = 2.49$). Henceforth, we will refer to the first of these pinch-off events as `back pinch-off', and to the second as `interior pinch-off'; the times for these events are denoted as $\tilde{t}_{p-o,bk}$ and $\tilde{t}_{p-o,int}$, respectively. These phenomena of successive back and interior pinch-offs in surfactant-free cases are in agreement with the numerical observations of \cite{Andredaki2021AcceleratingRegimes}.

The post second pinch-off dynamics are driven by rich rotation mechanisms around both the satellite and primary drops. The two pinch-off events create a small contracting liquid filament, characterised by local $Oh \sim 0.05$ and $L_{ligament}/R_{ligament} \sim 2.62$ and similar to those studied by \cite{Notz2004DynamicsFilament, Driessen2013StabilityFilaments}, and others. This small aspect ratio, in conjunction with the strong contracting kinetic energy derived from the second pinch-off, overcome capillary pressure forces to avoid an additional pinch-off instance in the small ligament. Observing the vorticity contours and the highlights at $\tilde{t} = 2.39 - 2.49$, an example of a vortex ring detachment can be noticed at the forming neck of the filament, where the vorticity boundary layer detaches and advects towards the centre of the bulbous region, pushing flow away from the neck and promoting the escape from pinch-off \citep{Hoepffner2013RecoilRing,Wang2019AFilaments,Constante-Amores2020DynamicsNumber}. As seen in $\tilde{t} = 2.57$, succeeding the small liquid filament retraction in the counter-flow direction ($\tilde{t} = 2.49$) we spot a counter expansion (see positive $\tilde{\omega}_{z}$ at $\tilde{t} = 2.57$) that generates the bullet-like satellite drop depicted at $\tilde{t} = 2.63$ and that eventually coalesces with the primary drop. The process of incorporation of the satellite into the primary drop exhibits the progressive growth of the finite liquid bridge \citep{Eggers1999CoalescenceDrops}, driven by the two opposite-signed high $|\tilde{\omega}_{x}|$ rings in its vicinity (see $\tilde{t} = 3.10$) as well as complex dynamics of interfacial waves across the newly-formed drop (see $\tilde{t} = 3.19 - 3.37$).

\subsection{Influence of surfactant parameters on cavity formation and encapsulation dynamics} \label{subseq:results_surfactants}

Having examined in detail the different encapsulation phenomena observed for high $Ca$ in surfactant-free interfaces, we continue our analysis to explore the influence of key surfactant dimensionless groups. We begin by highlighting the influence of Marangoni stresses on the cavity shape, followed by describing the relative effect of $\beta_{s}$ and $Bi$. Thereafter, we analyse the competition between the rates of interface surfactant adsorption and desorption, represented by $k$ and $Da$ simultaneously, as well as $Re$.

\subsubsection{Marangoni stresses, surface elasticity, and Biot number}

\begin{figure}
  \centerline{\includegraphics[scale=0.095]{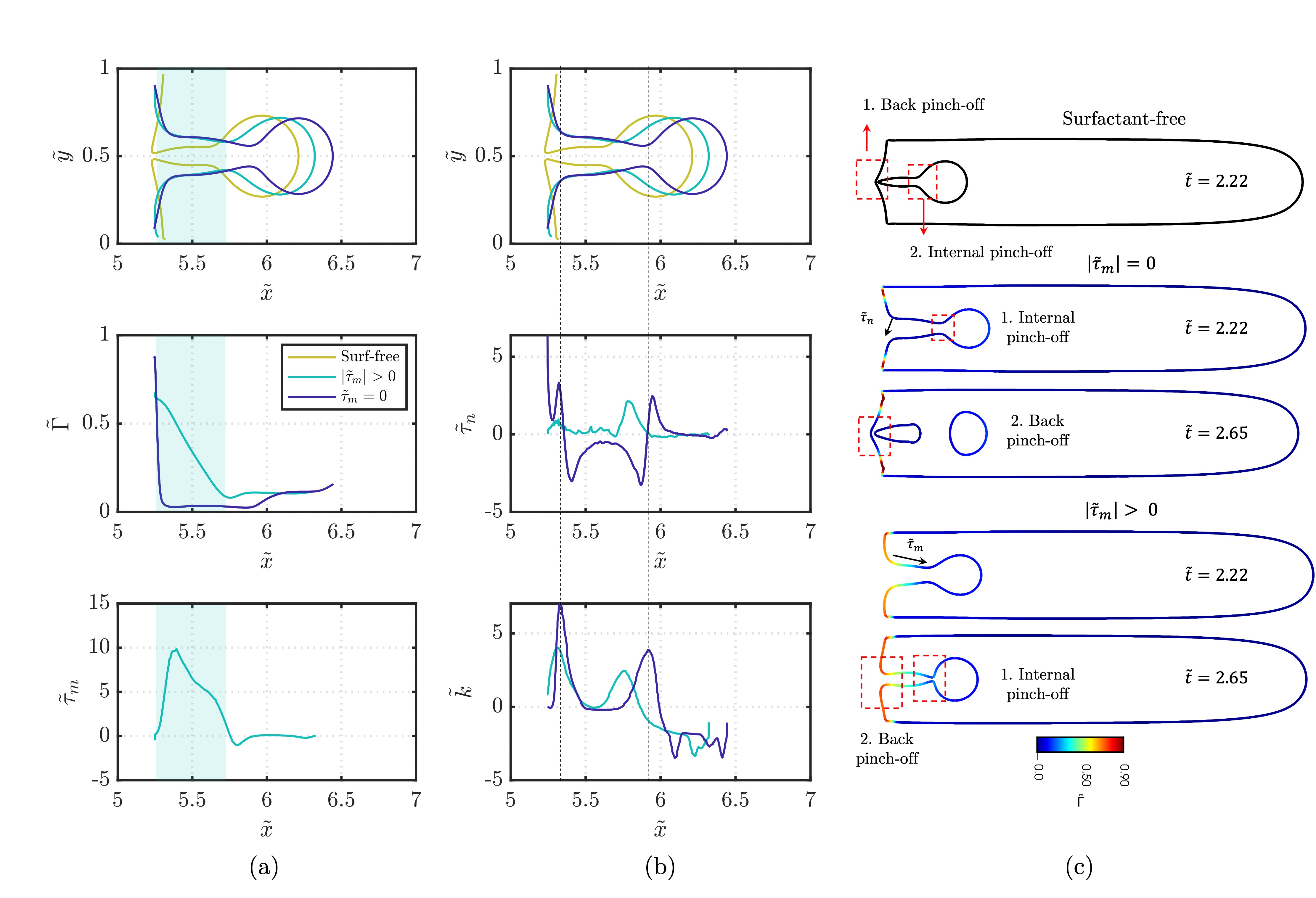}}
  \caption{Effect of Marangoni stresses on cavity formation and encapsulation. (a) Cavity shape (top), surfactant interfacial concentration (middle), and Marangoni stresses (bottom). (b) Cavity shape (top), normal stresses (middle), and interface curvature (bottom). $|\tilde{\tau}_{m}| > 0$ corresponds to Marangoni-supported, while $|\tilde{\tau}_{m}| = 0$ to Marangoni-free. (c) Schematic of pinch-off mechanisms. Plots on two-dimensional projection in x-y plane ($\tilde{z} = 0.5$) and at $\tilde{t} = 2.22$ for (a)-(b). $\tilde{t}_{p-o,bk} = 2.28, 2.45, 2.73$ and $\tilde{t}_{p-o,int} = 2.45, 2.66, 3.02$ for surfactant-free, $|\tilde{\tau}_{m}| = 0$, and $|\tilde{\tau}_{m}| > 0$, respectively. All other parameters remain unchanged from those specified in §\ref{subsec:sim_setup} for the base case.}
  \label{fig_interface_marangoni}
\end{figure}

To begin our analysis, we elucidate the behaviour of a surfactant-laden system under the conditions ascribed to the `base' case (refer to §\ref{subsec:sim_setup}) and compare it to that of surfactant-free interfaces. In order to isolate the effects of lower surface tension from those arising from Marangoni stresses on the encapsulation dynamics, we have set up an additional case, denoted by $|\tilde{\tau}_{m}| = 0$, in which we have suppressed the last term on the RHS of Eq. (\ref{eq:dim_momentum}) to
inhibit Marangoni stresses while simultaneously allowing the reduction of surface tension. Figures \ref{fig_interface_marangoni}(a)-(b) show the shape of the liquid cavity and a few other characteristics along its interface at $\tilde{t} = 2.22$. For the sake of clarity, we have duplicated the plots of interfacial shape on both column panels. Examination of the plots uncovers an almost surfactant-free interface at the liquid cavity for $|\tilde{\tau}_{m}| = 0$, which exhibits a significant drop in $\tilde{\Gamma}$ going from the edge to the interior of the cavity in the stream-wise direction and a slight surge after surpassing the incipient pinching neck. This rapid loss of surfactant for $|\tilde{\tau}_{m}| = 0$, as opposed to the Marangoni-supported case, $|\tilde{\tau}_{m}| > 0$, is explained by noticing that the operating conditions of the base case promote surfactant desorption (large $Da$ and small $k$) as $\tilde{\Gamma} \to 1 $ (see Eq. (\ref{eq:bulksource_nd})) as well as slow interface surfactant diffusion (large $Pe_{s}$). Reverting back to the remarks by \cite{Batchvarov2020EffectNumber,Atasi2018InfluenceMicrochannels}, the motion of the bubble prompts surfactant accumulation at the back due to drag forces from the liquid phase, partially counteracted by Marangoni stresses in the opposite direction for $|\tilde{\tau}_{m}| > 0$. The absence of Marangoni stresses in the $|\tilde{\tau}_{m}| = 0$ case leads to a continuous process of local surfactant accumulation at the back and desorption into the bulk, depleting the interface of surfactants.

Figure \ref{fig_interface_marangoni}(c) portrays a key difference in the encapsulation behaviour between the three cases, where the succesion of back pinch-off first and interior pinch-off second observed for clean interfaces is reversed in the presence of surfactants and the overall pinch-off dynamics are delayed (see the figure caption for the pinch-off times). This delay is accentuated for $|\tilde{\tau}_{m}| > 0$, as can be observed from the local peaks in $\tilde{\kappa}$ and $\tilde{\tau}_{n}$ of figure \ref{fig_interface_marangoni}(b), which, in the case of $|\tilde{\tau}_{m}| = 0$, are already pulling the interface towards the channel centre, signalling the onset of capillary neck closure. Part of the role of Marangoni stresses can be seen in the green-shaded region of figure \ref{fig_interface_marangoni}(a), where these tangential stresses counteract capillary forces at both the back and interior pinch-off regions. These observations are in complete agreement with the conclusions of \cite{Kamat2020Surfactant-drivenFilaments, Kamat2018RoleCascades,Constante-Amores2020DynamicsNumber} for liquid threads, although we note that in our system the Marangoni forces exerted on the interface are not sufficient to escape pinch-off completely. The inversion of back and interior pinch-off times in contaminated interfaces can be attributed to the noticeably non-uniform surfactant distribution along the liquid cavity, which displays its highest values in the regions neighbouring the capillary neck at the back of the cavity, lowering surface tension and 
providing additional disruption to neck closure in comparison to the interior cavity neck.

\begin{figure}
  \centerline{\includegraphics[scale=0.09]{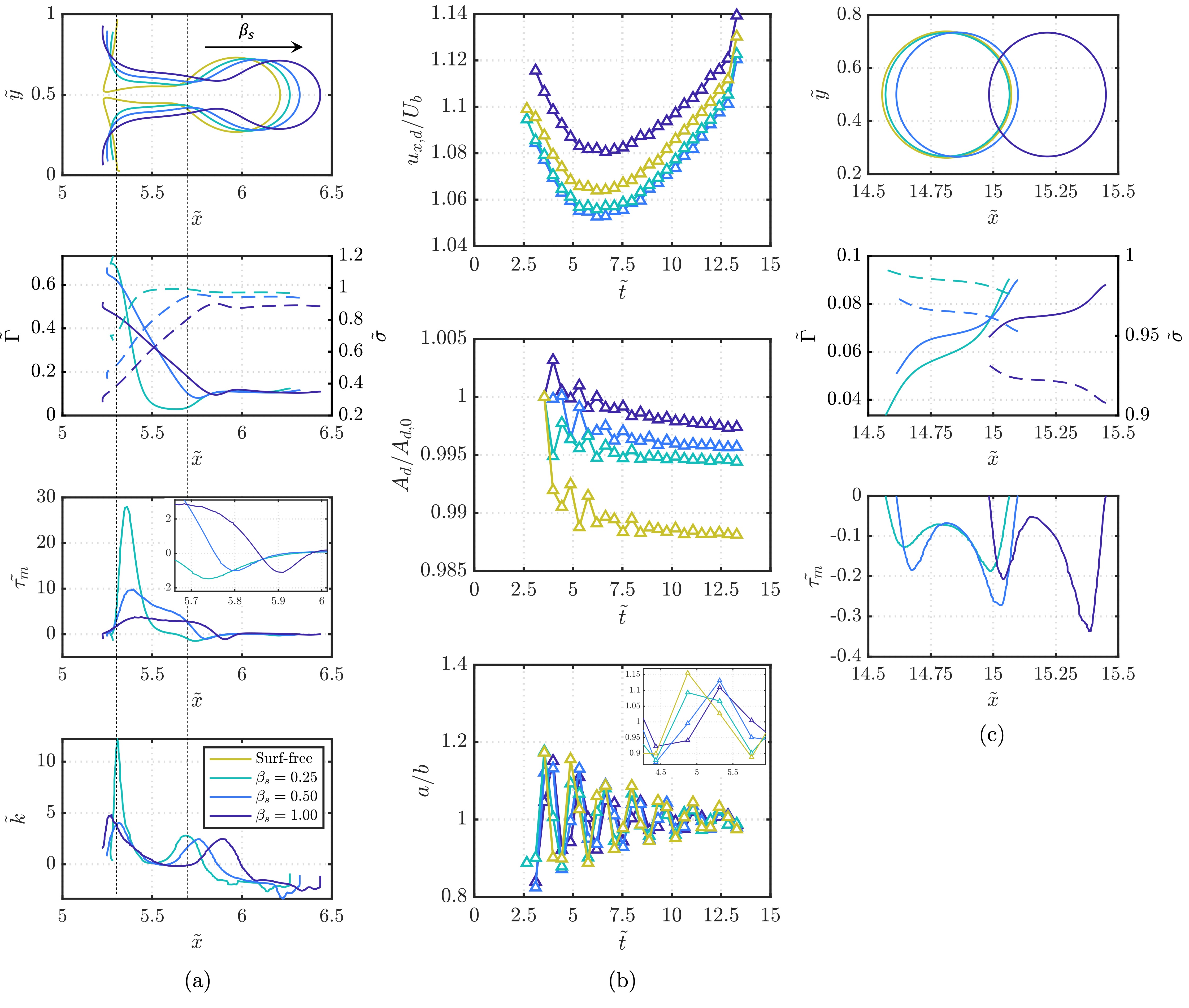}}
  \caption{Effect of $\beta_{s}$ on cavity formation and encapsulation. (a) Cavity shape, surfactant interfacial concentration (left axis, continuous line) and surface tension (right axis, dotted line), Marangoni stresses, and interface curvature at $\tilde{t} = 2.22$. (b) Temporal evolution of primary encapsulated drop velocity normalised by bubble velocity, primary drop interfacial area normalised by area after pinch-off, and ratio between drop length in the $x$ and $y$ directions. Data points computed for all cases every $10^{-3}s$. (c) Primary encapsulated drop shape, surfactant interfacial concentration (left axis, continuous line), and surface tension (right axis, dotted line) and Marangoni stresses at $\tilde{t} = 7.82$. Plots on two-dimensional projection in x-y plane ($\tilde{z} = 0.5$) for (a,c). $\tilde{t}_{p-o,bk} = 2.28,2.52,2.73,2.90$ and $\tilde{t}_{p-o,int} = 2.45,2.64,3.02,3.94$ for surfactant-free, and $\beta_{s} = 0.25,0.50,1$, respectively. All other parameters remain unchanged from those specified in §\ref{subsec:sim_setup} for the base case}
 \label{fig_interface_beta}
\end{figure}

For the first part of our parametric investigation, we study the effect of $\beta_{s}$ on encapsulation and the post pinch-off dynamics. As figure \ref{fig_interface_beta}(a) depicts, and in line with the studies of \cite{Antonopoulou2021EffectPrinting,Zhong2018AnalysisNozzle} in the context of ink-jet printing, increasing surfactant strength gradually increases the speed of cavity formation and decreases the rate of neck thinning at both the back and interior of the cavity (see locations of cavity nose in figure \ref{fig_interface_beta}(a) and pinch-off times in the figure caption). This delay is materialised in the larger $|\tilde{\kappa}|$ peaks seen for $\beta_{s} = 0.25$ at the incipient pinching necks, as well as the higher $|\tilde{\tau}_{m}|$ that arise at their vicinity, seeking to counteract capillary draining \citep{Ambravaneswaran1999EffectsBridges}. Figure \ref{fig_interface_beta}(b) illustrates the progression of the primary encapsulated drop as it travels across the bubble domain in the form of its volume-averaged $x$-velocity, $u_{x,d}$, normalised by $U_{b}$ (top), interfacial area, $A_{d}$, normalised by its area after pinch-off, $A_{d,0}$ (middle), and ratio between drop length in the $x$ and $y$-directions, $a/b$ (bottom). The encapsulated drops exhibit three common characteristics irrespective of the presence of surfactants; namely, a higher stream-wise velocity than the bubble nose, highlighting the potential for a scenario in which the bubble nose interface is ruptured by the drop, an overall decrease in interfacial area, and dampening $a/b$ oscillations in time, which culminate in virtually spherical drops. This bubble nose rupture is further analysed in §\ref{subseq:results_deformation}. Under the conditions considered, the addition of a strong surfactant ($\beta_{s} = 1$) decreases the rate of surface area reduction and significantly increases drop velocity relative to the bubble nose. The non-monotonic behaviour of $u_{x,d}/U_{b}$ with regard to a clean interface and weak surfactants ($\beta_{s} = 0.25 - 0.50$) can be explained by observing in figure \ref{fig_interface_beta}(c) the spatial distribution of $\tilde{\Gamma}$, $\tilde{\sigma}$ and $\tilde{\tau}_{m}$ in the encapsulated drop at $\tilde{t} = 7.82$. Despite the largest $|\tilde{\tau_{m}}|$ opposing drop propagation in the stream-wise direction for the strongest surfactant, it is seen that these effects are fully countered by the lower surface tension across the drop's domain, yielding a higher drop velocity than the surfactant-free case. In contrast, the weak surfactant cases feature an almost clean interface with surface tension values very close to the clean case, but with non-zero Marangoni effects that induce a delay in drop propagation.

The morphological deformations that characterise drop behaviour after pinch-off, as depicted in figure \ref{fig_interface_beta}(b)(bottom), demonstrate the presence of a semi-periodic oscillating pattern monotonically influenced by surface elasticity. In the inset of this figure, a higher deviation from a fully spherical drop can be appreciated for the surfactant-free case in both the major crests and troughs of the oscillatory structures. Considering these results, we draw parallels with the study of \cite{Wang2019DeformationSystem} in an ink-jet system, whose main results suggest that the extent of deformation decreases with $Ca$ (after a critical $Ca$ has been reached), in line with our numerical observations for increasing $\beta_{s}$. Estimating the frequency of $a/b$ oscillations through the classic expression of \cite{Rayleigh1879OnJets} ($f \sim \sqrt{\sigma/\rho_{c}r_{d}^{3}}$, where $r_{d}$ is the drop's spherical equivalent radius) at the latest time recorded ($\tilde{t} = 13.29$), we obtain $\tilde{f} \sim $ $1.49, 1.55, 1.55, 1.57$ for surfactant-free and $\beta_{s} = 0.25, 0.50, 1$, respectively, implying a dominating effect of lower drop radius over lower surface tension on the frequency of deformation cycles for increasing $\beta_{s}$.

\begin{figure}
  \centerline{\includegraphics[scale=0.095]{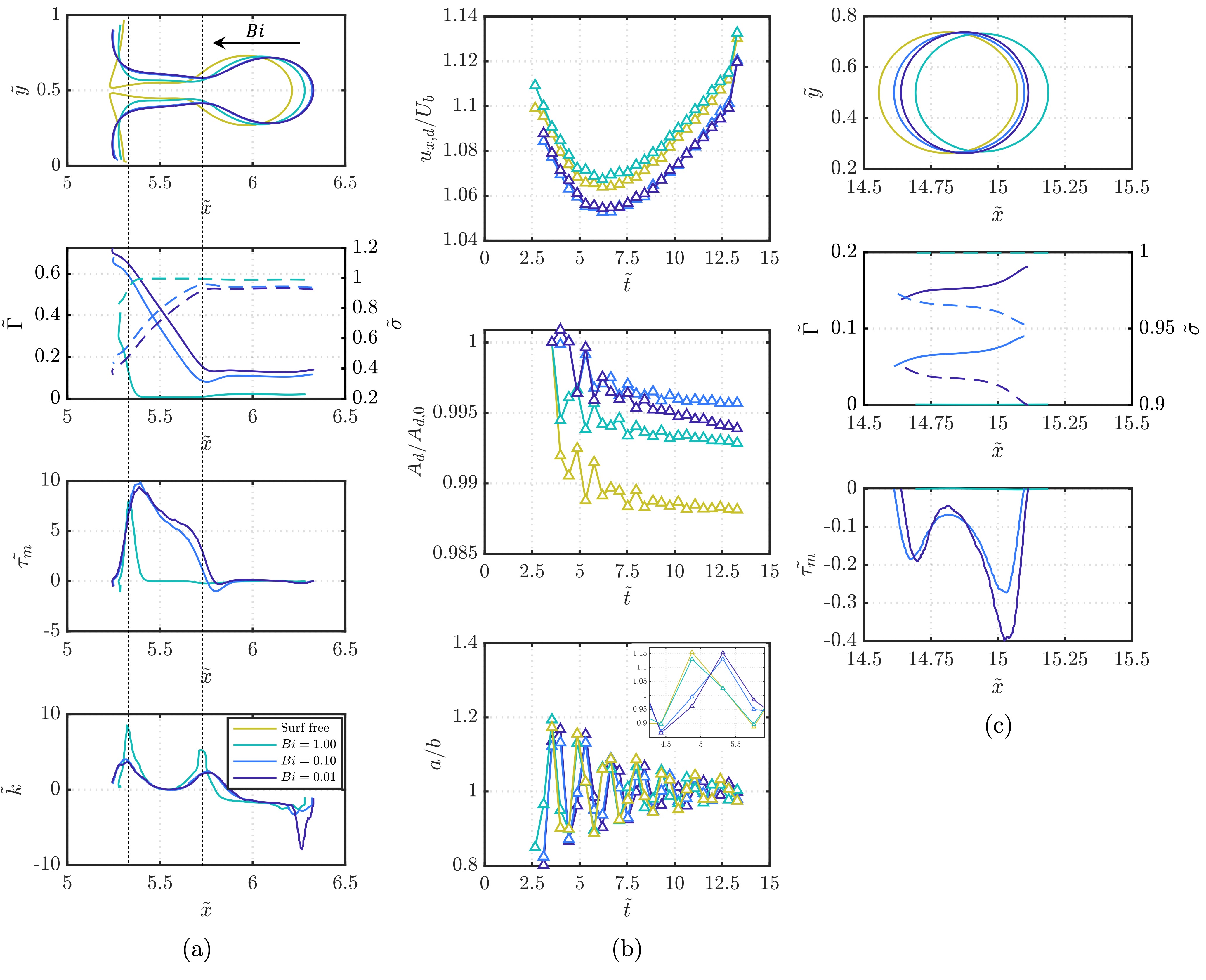}}
  \caption{Effect of $Bi$ on cavity formation and encapsulation. (a) Cavity shape, surfactant interfacial concentration (left axis, continuous line) and surface tension (right axis, dashed line), Marangoni stresses, and interface curvature at $\tilde{t} = 2.22$. (b) Temporal evolution of primary encapsulated drop velocity normalised by bubble velocity, primary drop interfacial area normalised by area after pinch-off, and ratio between drop length in the $x$ and $y$ directions. Data points computed for all cases every $10^{-3}s$. (c) Primary encapsulated drop shape, surfactant interfacial concentration (left axis, continuous line), surface tension (right axis, dashed line), and Marangoni stresses at $\tilde{t} = 7.82$. Plots on two-dimensional projection in $x-y$ plane ($\tilde{z} = 0.5$) for (a,c). $\tilde{t}_{p-o,bk} = 2.28, 2.48, 2.73, 2.80$ and $\tilde{t}_{p-o,int} = 2.45, 2.75, 3.02, 3.21$ for surfactant-free and $Bi = 1, 0.10, 0.01$, respectively. All other parameters remain unchanged from those specified in §\ref{subsec:sim_setup} for the base case}
 \label{fig_interface_bi}
\end{figure}

The effects arising from altering $Bi$ in the range $0.01-1$ are depicted in the plots of figure \ref{fig_interface_bi}. An interface covered by a highly soluble surfactant (i.e., high $Bi$) undergoes rapid rates of mass exchange between the interface and bulk as a result of its comparatively low (high) characteristic desorptive (inertial) time. This is demonstrated by noticing the significantly lower $\tilde{\Gamma}$ distributed along the liquid cavity for the highest $Bi$ considered, which, barring the surfactant-free case ($Bi \to \infty$), exhibits the slowest rate of cavity formation and the fastest pinch-off events (see location of cavity nose in figure \ref{fig_interface_bi}(a)(top) and figure caption for the pinch-off times), in alignment with preceding investigations (see \cite{Milliken1994TheSolubility}). Similar to our previous remarks, the signature $|\tilde{\tau}_{m}|$, $|\tilde{\kappa}|$, and capillary pressure peaks (not shown) that signal the onset of capillary neck closure and the advent of pinch-off are already developing in a noticeable manner for $Bi = 1$, in contrast to $Bi = 0.01 - 0.10$.

The post pinch-off temporal evolution of the primary drop, as depicted in figure \ref{fig_interface_bi}(b), reveals similar trends to those referenced in the above analysis of $\beta_{s}$ in terms of drop shrinkage, its attenuation in surfactant-laden cases, and semi-periodic drop size oscillations. Nonetheless, we observe the incidence of non-monotonic dynamics with regard to $Bi$ and the surfactant-free case. In particular, the highest values of $u_{x,d}/U_{b}$ appear to occur for the most soluble surfactant, followed by the surfactant-free case, leaving the less soluble cases with the lowest $u_{x,d}/U_{b}$. The disruption of the direct relationship between higher pinch-off times and higher $u_{x,d}/U_{b}$ seen for $\beta_{s}$ suggests a complex interplay between Marangoni stresses, surface tension, pinch-off times, and drop velocity with varying $Bi$. The spatial locations of the encapsulated drops at $\tilde{t} = 7.82$ (figure \ref{fig_interface_bi}(c) (top)) uncover an additional layer of non-monotonic behaviour, where the overall fastest (slowest) drop corresponds to $Bi = 1$ (surfactant-free), despite these two cases being the closest in terms of pinch-off times and local $\tilde{\sigma}$. The large velocities exhibited by $Bi = 1$ in comparison to lower $Bi$ and the surfactant-free case is explained by noticing its virtually clean interface and thus the complete absence of Marangoni stresses to counteract its axial motion at later times, while also having a low enough $\tilde{\sigma}$ at the earlier times to increase the rate of cavity formation and delay pinch-off.

\subsubsection{Effect of surfactant adsorption/desorption kinetics ($Da$ and $k$) and $Re$}

\begin{figure}
  \centerline{\includegraphics[scale=0.1]{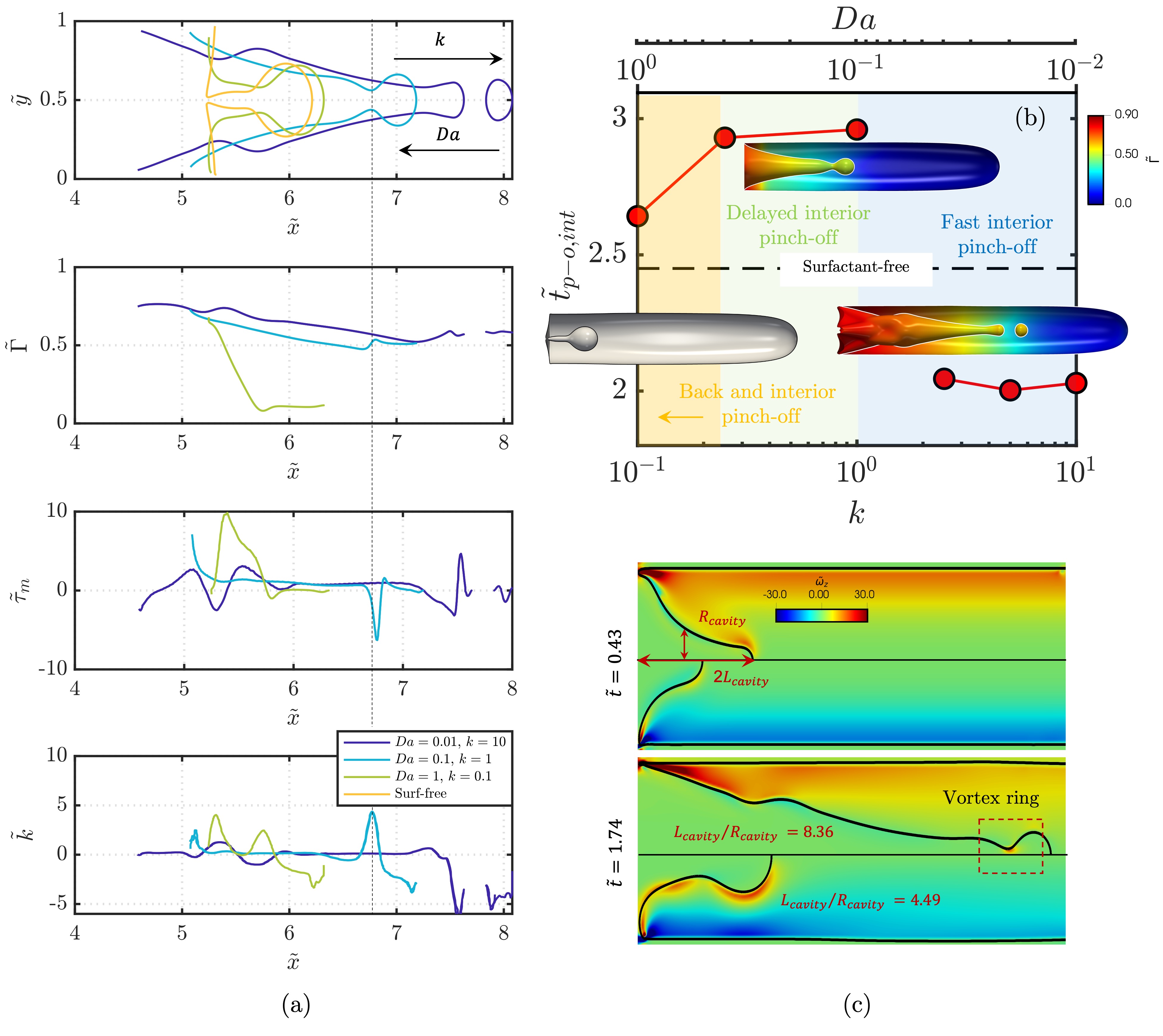}}
  \caption{Effect of $k$ and $Da$ on cavity formation and encapsulation. (a) Cavity shape, surfactant interfacial concentration, Marangoni stresses, and interface curvature at $\tilde{t} = 2.22$. (b) Interior pinch-off time vs. $k$ and $Da$, and schematic of the three regions that divide thinning behaviour. (c) Contour plots of normalised vorticity in z-direction and comparison of cavity shape for $k = 0.1, Da = 1.0$ (top) and $k = 10.0, Da = 0.01$ (bottom). The values of $R_{cavity}$ were calculated at the midpoint between the cavity nose and the start of the cavity. Plots on two-dimensional projection in x-y plane ($\tilde{z} = 0.5$). $\tilde{t}_{p-o,int} = 2.45, 2.64, 2.93, 2.96, 2.05, 2.00, 2.03$ for surfactant-free and $k = 0.10, 0.25, 1, 2.50, 5.00, 10.00$ ($Da = 1, 0.40, 0.10, 0.04, 0.02, 0.01$), respectively. A few cases of the full range of $k$ have been omitted in (a) for an easier visualisation. All other parameters remain unchanged from those specified in §\ref{subsec:sim_setup} for the base case}
 \label{fig_interface_da}
\end{figure}

We now proceed to examine the influence of directly contrasting the characteristic times of adsorption and desorption, along with the initial interface saturation through varying $k$ and $Da$ simultaneously. This concurrent variation allows us to maintain all other parameters constant. Figure \ref{fig_interface_da}(a) portrays a monotonic response of the cavity depth and $\tilde{\Gamma}$ with $k$ and $Da$, where the rapid rates of surfactant adsorption inherent to the high $k$ (low $Da$) cases are materialised in the relatively large and comparatively constant $\tilde{\Gamma}$ distributions along the cavity. In line with our previous results, general reductions in local surface tension at the bubble rear, brought about by higher $k$ or $\beta_{s}$ and lower $Bi$ conditions, promote a faster and deeper infiltration of liquid into the bubble domain. In figure \ref{fig_interface_da}(b) we record the time until the first pinch-off instance at the bubble interior as a function of $k$ for the entire parameter space explored. A retardation of the capillary instabilities that lead to end-pinching is seen as $k$ is increased in the range $0$ (surfactant-free) $< k < 1$, consistent with our previous remarks about the combined effect of $\tilde{\sigma}$ and Marangoni stresses on opposing drop encapsulation. This behaviour is mirrored by the $\tilde{\tau}_{m}$ and $\tilde{\kappa}$ peaks exposed in the vicinity of the nascent pinching necks for the lowest $k$ cases displayed in figure \ref{fig_interface_da}(a) ($k = 0.1 - 1.0$), while the oscillations seen for $k = 10$ correspond to capillary waves in the liquid cavity, as reported by \cite{Constante-Amores2020DynamicsNumber} for long liquid ligaments.

Similar to our previous analyses of $\beta_{s}$, $Bi$, and Marangoni-free case for $k = 0.10$, it is found that increasing $k$ in the range $k < 0.10$ results in a sequence inversion of the two main pinch-off events ($\tilde{t}_{p-o,bk}$ $>$ $\tilde{t}_{p-o,int}$) identified in clean interfaces ($\tilde{t}_{p-o,bk}$ $<$ $\tilde{t}_{p-o,int}$, see caption of figure \ref{fig_interface_da}). Interestingly, it is also found that for all cases above $k = 0.25$, the large accumulation of surfactants nearby the bubble rear provides a strong resistance to capillary pressure buildup in those regions, resulting in the partial elimination of the `back pinch-off' event until bubble bursting events occur. This will be further discussed in the following section. Another notable feature to highlight from figure \ref{fig_interface_da}(b) is the sharp decrease in $\tilde{t}_{p-o,int}$ above the critical value of $k = 1$. This  discontinuity in the function $\tilde{t}_{p-o,int}$ vs. $k$, in conjunction with our observations about the elimination of the back pinch-off suggest a partition of the encapsulation dynamics into two major regions according to the rates of adsorption/desorption: i) the above described region for $k < 1$, in which the presence of surfactants exerts a stabilising effect, leading to a delay in neck thinning in relation to a fully clean interface; and, ii) a region of rapid cavity formation in the axial direction alongside a significant acceleration of the first interior pinch-off event for $k > 1$. The first region is further segregated by the occurrence ($k < 0.25$, region i.a) or elimination ($k > 0.25$, region i.b) of the back pinch-off. An explanation for the marked transition from the first to the second region 
in terms of thinning rates is as follows. A general increase of surfactant coverage due to high rates of adsorption introduces high rates of viscous and inertial deformation to the interface, as noted by \cite{Bazhlekov2006NumericalFlow}, hence, increasing the axial velocity of the cavity and its aspect ratio. 
Recalling the observations of \cite{Wang2019AFilaments} and \cite{Ambravaneswaran1999EffectsBridges} in contacting/expanding liquid filaments, we show in figure \ref{fig_interface_da}(c) how in cases above the critical $k$, $L_{cavity}/R_{cavity}$ surpasses a given threshold value to enter the beak-up regime and accelerate end-pinching, even if the local values of surface tension are lower than those of smaller $k$ conditions.

\begin{figure}
  \centerline{\includegraphics[scale=0.075]{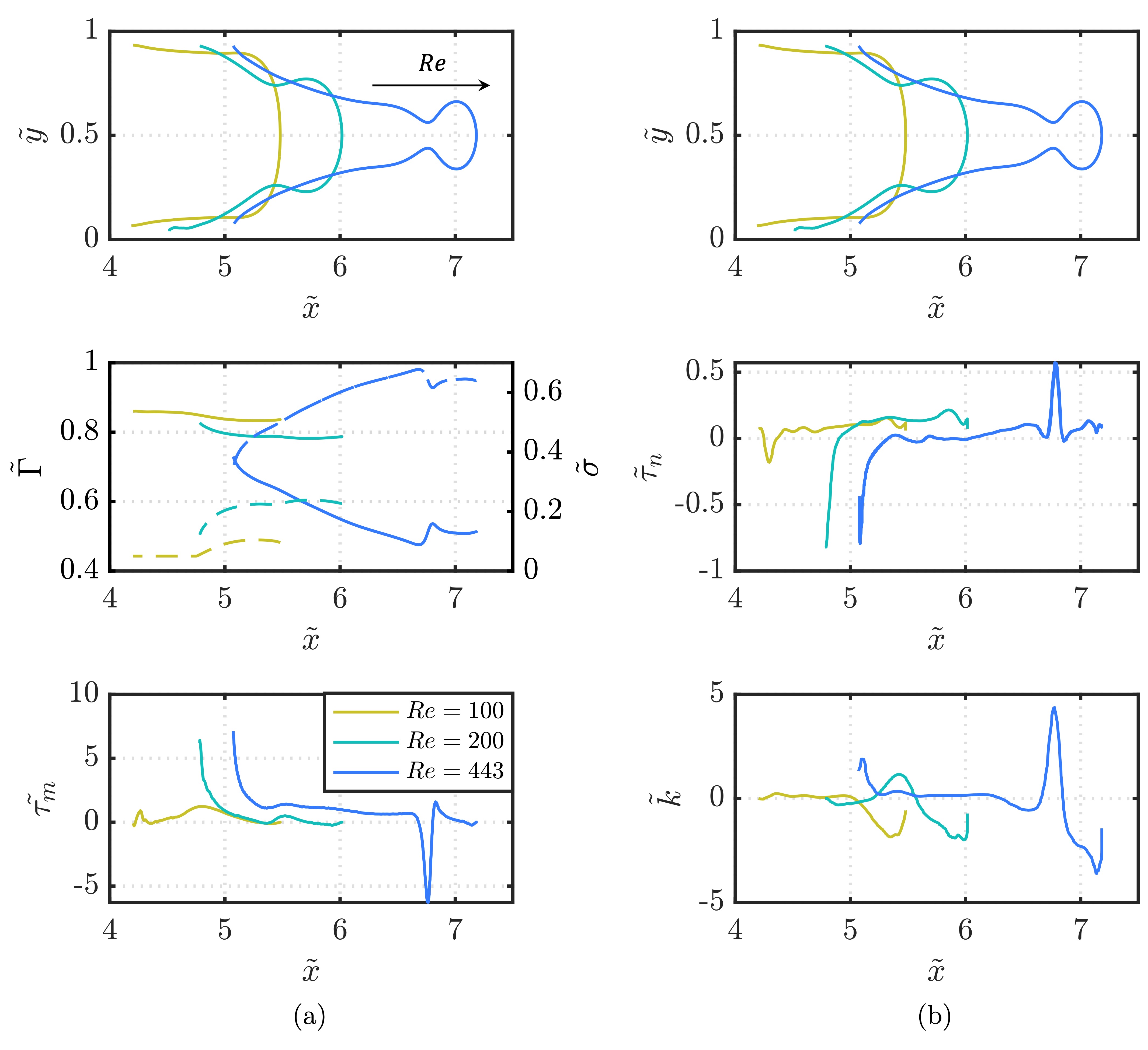}}
  \caption{Effect of $Re$ on cavity formation and encapsulation. (a) Cavity shape (top), surfactant interfacial concentration (middle, left axis, continuous line) and surface tension (middle, right axis, dotted line), and Marangoni stresses (bottom). (b) Cavity shape (top), normal stresses (middle), and interface curvature (bottom). Plots on two-dimensional projection in x-y plane ($\tilde{z} = 0.5$) and at $\tilde{t} = 2.22$. For the three cases, $Ca = 0.0693$, $Pe_{c,s} = 100$, $Bi = 0.10$, $Da = 0.10$, $k = 1$, and $\beta_{s} = 0.50$. $\tilde{t}_{p-o,int} = 2.96$ for $Re = 443$. Pinch-off not observed for $Re = 100 - 200$. For this set of results, $Da = 0.10$ and $k = 1$. All other parameters remain unchanged from those specified in §\ref{subsec:sim_setup} for the base case}
 \label{fig_interface_re}
\end{figure}

Moving on to the final exploration of the dimensionless space of the system, we test the influence of varying $Re$ in the range $100 - 443$ ($We = 6.93 - 30.70$) and maintaining all other characterising numbers of Eq. (\ref{eq:dimless}) constant: $Ca = 0.0693$, $Pe_{c,s} = 100$, $Bi = 0.1$, $Da = 0.1$, $k = 1$, and $\beta_{s} = 0.5$. Figure \ref{fig_interface_re} demonstrates the pronounced effects of $Re$ on cavity formation and the overall surfactant dynamics, where the presence of comparatively lower inertial forces of reduced $Re$ and constant $Ca$ (lower $We$) appears to oppose the complete development of the liquid cavity and the process of drop entertainment, whilst still allowing the rear curvature inversion described for large $Ca$ (see §\ref{subseq:results_clean}). The spatial profiles of $\tilde{\Gamma}$ show a cavity largely covered in a uniform manner by surfactants for $Re = 100$, manifested in the very low $\tilde{\tau}_{m}$ exerted and the virtually planar $\tilde{\kappa}$ of a significant portion of its cavity. This contrasting behaviour against the highly non-uniform and smaller $\tilde{\Gamma}$ describing high $Re$ provides evidence for arguing that, in the locality of the cavity, inertio-capillary forces are greatly overcome by viscous-related forces (higher $Oh$), strongly opposing liquid infiltration, necking, and ultimately pinching. We therefore highlight the aggregate effect of both $Ca$ and $Re$ on the successful entrapment of liquid drops within the bubble domain, which necessitates an initial bubble rear inversion, a sufficiently fast cavity development, and the incidence of capillary pressure buildup that leads to pinch-off.

\subsection{Bubble bursting, back deformations, and regime maps} \label{subseq:results_deformation}

\begin{figure}
  \centerline{\includegraphics[scale=0.055]{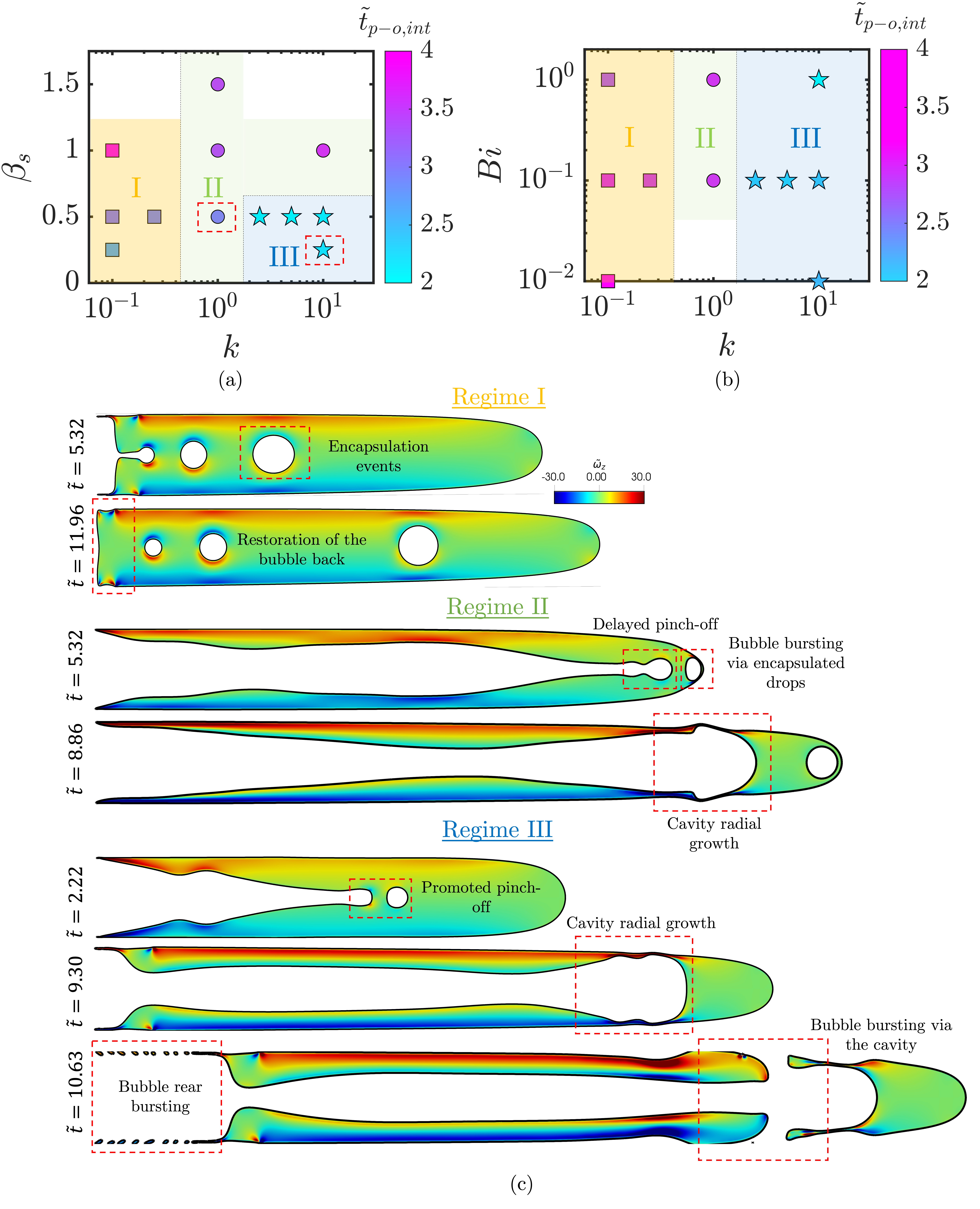}}
  \caption{Encapsulation and bursting regime maps in the (a) $\beta_{s}$-$k$ and (b) $Bi$-$k$ spaces. $Bi = 0.10$ for (a) and $\beta_{s} = 0.50$ for (b). The pinch-off times reported correspond to those of the first interior pinch-off event observed. (c) Contour plots of normalised vorticity in z-direction and comparison for the three regimes identified. All other parameters remain unchanged from those specified in §\ref{subsec:sim_setup} for the base case}
 \label{fig_map_beta_Bi_k}
\end{figure}

We finalise our discussion by providing a more detailed account of the post pinch-off dynamics, focusing specifically on the process of bubble bursting via the liquid cavity/entrapped drops and the temporal fate of the bubble back and accompanying trailing structures. In figure \ref{fig_map_beta_Bi_k}(a)-(b) we summarise the three main bursting behaviours observed in the $\beta_{s}$-$k$ and $Bi$-$k$ ($Da$) spaces, leaving all other conditions specified in the baseline case unchanged. As captured in the figure, a first regime (I), distinctly bounded by a surfactant-free interface and $k$ below the critical value defined in the previous section, is identified for all $\beta_{s}$ and $Bi$ tested. This regime adheres to a behaviour mode of slow infiltration of liquid, monotonic delay of the first interior pinch-off event with increasing (decreasing) $\beta_{s}$ ($Bi$), and most importantly, eventual restoration of the interfacial morphology of the bubble back, depicted in figure \ref{fig_map_beta_Bi_k}(c, top). This restoration ensures a pseudo-stable entrapment of one or multiple drops of varying size and $\tilde{\Gamma}$ within the bubble domain, only disrupted by a potential rupture of the bubble nose by the first drop. We note that this rupture was not observed for any of the cases encompassed within (I) given the channel length employed in our simulations and the significant elongation of the bubble across the channel. The stability of the encapsulated drops-bubble compound in regime I poses a potential avenue for future research.

A second regime (II) emerges as $k$ approaches the critical value, where a much faster and deeper liquid cavity develops across the bubble and, ensuing from diminished capillary forces, further delays to pinch-off are recorded. These delays lead to the entrapment of the first drop in regions approaching the bubble nose in the axial direction, promptly leading to its rupture, as illustrated at $\tilde{t} = 5.32$ in figure \ref{fig_map_beta_Bi_k}(c, middle). Following a series of end-pinching events, the liquid cavity undergoes several instances of escapes of pinch-off, in conjunction with a surge of surface capillary waves across its domain, and a radial growth that gives rise to a large bulbous end (see $\tilde{t} = 8.86$). These mechanisms are closely related to the intricate phenomena stemming from retracting liquid filaments, as addressed in \cite{Anthony2019DynamicsFilaments,Constante-Amores2020DynamicsNumber}. This substantial growth causes the liquid cavity to swiftly burst the interface near the channel walls, thereby fully separating the original bubble into two individual entities. Notably, the fate of the newly-formed small entity located further down the stream-wise direction mirrors that of the original bubble as new instances of encapsulation are observed. These bursting mechanisms are wholly in line with the observations of \cite{Atasi2018InfluenceMicrochannels} for close-to-spherical bubbles under similar conditions of $k$ and $Ca$ (see figure 16 of this reference) and the reports of \cite{Izbassarov2016AContraction/expansion} for contraction/expansion channels with a viscoelastic liquid phase. Following this busting, capillary forces will attempt to minimise surface area by closing the openings of the newly-formed bubbles.

The third regime (III) is confined to regions of above critical $k$ with moderate $\beta_{s}$ and $0.01<Bi<1$ and closely resembles the mechanisms of II, with the pinch-off times being the key differentiating characteristic. In III, the fast expansion of the liquid cavity expedites the first pinch-off with respect to the surfactant-free case (see previous sub-section). Although the location of the first entrapment is much farther away from the bubble nose than in II, the encapsulated drops exhibit a comparatively faster velocity (e.g., $u_{x,d}/U_{b} = 1.27$ and $u_{x,d}/U_{b} = 1.47$ at $\tilde{t} = 4.87$ for the cases in II and III emphasised in figure \ref{fig_map_beta_Bi_k}(a), respectively), also leading to bubble nose ruptures by the drops and the consequent radial rupture by the bulbous end of the cavity mentioned above (see $\tilde{t} = 10.63$). We now briefly draw attention to the multitude of non-axisymmetric structures that unfold in II and III at the trailing ends of the original bubble after the rupturing sequence. Instead of the restoration of the bubble back seen in I, the liquid drag and prevailing adsorption-induced accumulation of surfactant promotes the thinning and elongation of the bubble back in the counter-flow direction, as highlighted in figure \ref{fig_map_beta_Bi_k}(c, bottom), whereupon rich interfacial dynamics transpire, including numerous bubble bursting and coalescence instances. These intricate dynamics at the trailing structures of the bubbles are consistent with the numerical results of \cite{Nath2017MigrationRegime} (e.g, figure 15 of this reference).


\section{Concluding remarks}\label{sec:conclusions}

This work presents an in-depth characterisation of the highly unsteady mechanisms describing elongated bubbles flowing across liquid-filled channels via numerical simulations. These mechanisms arise in opposition to the well-known steady-state phenomena of Taylor bubbles under the traditional assumptions of negligible inertial/viscous effects, symmetric cross-section channels, and the absence of surface-active agents. As reported in limited instances (see \cite{Atasi2018InfluenceMicrochannels} and \cite{Sauzade2013InitialOils}) and confirmed here, interesting dynamics emerge as $We\to$ \textit{O}(10). Notably, the concomitant reduced capillary effects of high $Ca$ (and $We$) induce a loss of sphericity at the bubble's front and back, triggering an elongation in the flow direction of the former and a flattening and subsequent curvature inversion of the latter. The underlying mechanisms by which this curvature inversion evolves into a liquid cavity infiltrating the bubble, how this cavity collapses into small liquid drops entrapped in the bubble domain, and the overall role of surfactants in these processes have been analysed in detail. We emphasise in this study the multiple parallels that can be readily drawn between the various interfacial features, singularities, and intricate  interplay amongst the multitude of physical mechanisms that ensue in ours and other well-known systems involving capillary liquid breakup, such as inkjet printing, two-phase microfluidics, and contracting liquid filaments.

By performing a systematic sweep of a number of surfactant parameters and characterising dimensionless groups, we have elucidated the coupled effect of lower surface tension and Marangoni stresses, brought about by the surfactants, and their interactions with inertia and viscosity. Two distinct modes of cavity breakup/drop encapsulation were found for comparatively low rates of surfactant adsorption (low $k$): one closure of the liquid cavity at the bubble's rear and one end-pinching breakup mode at the interior of the cavity. Our simulations have shown that the combined and individual effects of both surface tension and Marangoni stresses induce a delayed response in terms of pinch-off times, as well as an inversion of the sequence of breakup modes with respect to clean interfaces. Provided that $k$ is low enough, these effects are maintained for increasingly strong (large $\beta_{s}$) and less soluble (small $Bi$) surfactants. For these cases, complex post pinch-off non-monotonic behaviours were observed in terms of drop size, semi-periodic deformations across the bubble, and velocity relative to the bubble nose, suggesting noticeable effects from Marangoni stresses exerted on the encapsulated drops. Under the baseline conditions considered, a critical value of $k = 1$ represents a phenomenological limit from which an increase of $k$ represents an acceleration of the second mode of cavity breakup when compared to an uncontaminated interface due to the fast and deep liquid infiltration in the bubble, promoting capillary breakup. A summarising map with three well-defined regimes of behaviour was constructed in the $\beta_{s}$-$k$ and $Bi$-$k$ ($Da$) spaces, where regime I, limited by the critical $k$, allows for a semi-stable drop entrapment due to the eventual closure and restoration of the bubble back in the aftermath of pinch-off. Conversely, in regimes II (delayed pinch-off) and III (expedited pinch-off), the deep infiltration of the cavity that precedes end-pinching, together with the high relative velocity of the entrapped drops are responsible for the subsequent bursting of the bubble nose. These two regimes, in addition, display a multitude of rich interfacial events following the cavity end-pinching, including escapes from pinch-off, radial growth, and an eventual bubble bursting in the vicinity of the channel walls.

To the best of our knowledge, this is the first thorough and detailed characterisation of the phenomenon of drop encapsulation in moving bubbles in the presence of surfactants. There are, however, a few avenues worthy of further pursuit in this topic. For instance, an examination of the influence of surfactant parameters, $\beta_{s}$ for example, on the critical $k$ value. Based on our numerical results, it is likely that this critical $k$ will decrease with increasing $\beta_{s}$, but additional evidence is required. A deeper exploration of the stability of the encapsulated drops in regime I may also be of interest to determine if there exists a set of conditions that allow for a perpetual entrapment of these drops, and the potential applications of such systems in the context of emulsification, as remarked previously by \cite{Izbassarov2016AContraction/expansion}. \\

\textbf{Declaration of Interests}. The authors report no conflict of interest.\\ 

\textbf{Acknowledgements and funding.} This work is supported by the Engineering and Physical Sciences Research Council, United Kingdom, through the EPSRC MEMPHIS (EP/K003976/1) and PREMIERE (EP/T000414/1) Programme Grants. O.K.M. acknowledges funding from PETRONAS and the Royal Academy of Engineering for a Research Chair in Multiphase Fluid Dynamics. We acknowledge HPC facilities provided by the Research Computing Service (RCS) of Imperial College London for the computing time. D.J. and J.C. acknowledge support through HPC/AI computing time at the Institut du Developpement et des Ressources en Informatique Scientifique (IDRIS) of the Centre National de la Recherche Scientifique (CNRS), coordinated by GENCI (Grand Equipement National de Calcul Intensif) Grant 2022 A0122B06721. P.P. acknowledges the doctoral scholarship from the Colombian Ministry of Science, Technology and Innovation, MINCIENCIAS. The authors acknowledge with gratitude A. Batchvarov for inspiring this work and for all fruitful discussions. We also acknowledge helpful comments on the manuscript from A. Lavino. 

\appendix
\section{Mesh independence and resolution analysis}\label{ap:mesh}

\begin{figure}
     \centering
     \begin{subfigure}[b]{0.328\textwidth}
         \centering
         \includegraphics[width=\textwidth]{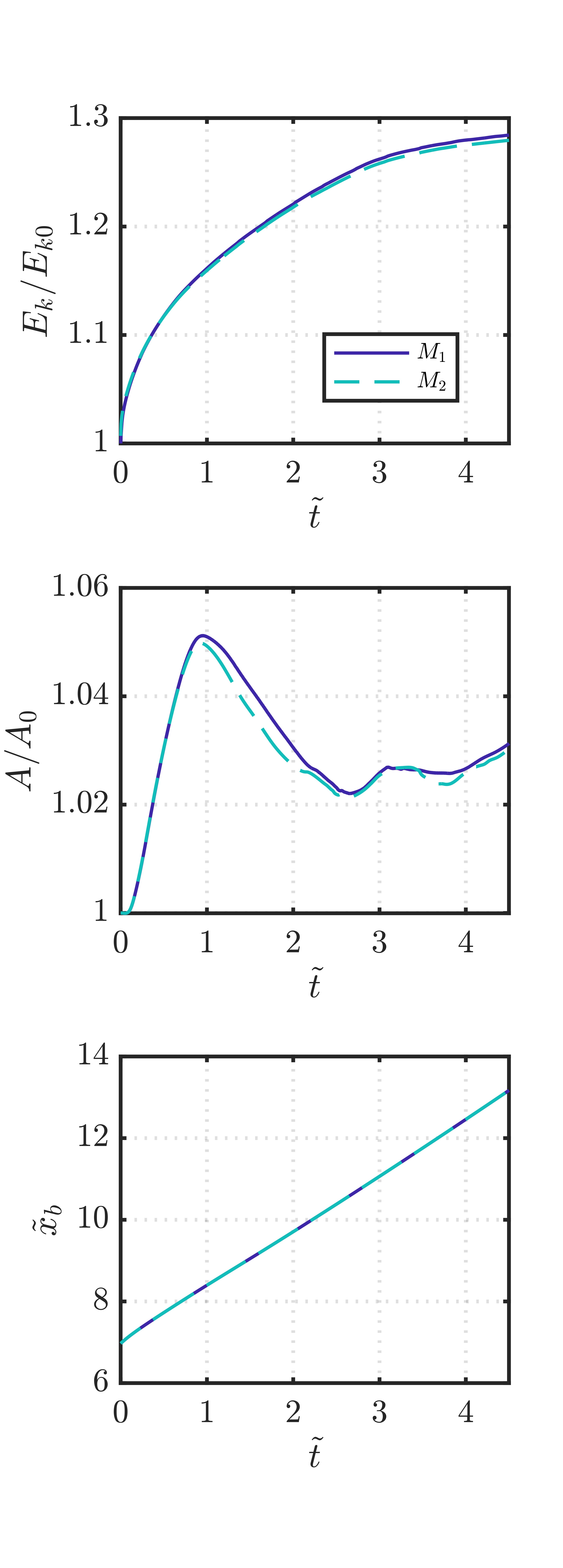}
         \caption{Case 0 (surfactant-free)}
         \label{fig:mesh_case0}
     \end{subfigure}
     \hfill
     \begin{subfigure}[b]{0.328\textwidth}
         \centering
         \includegraphics[width=\textwidth]{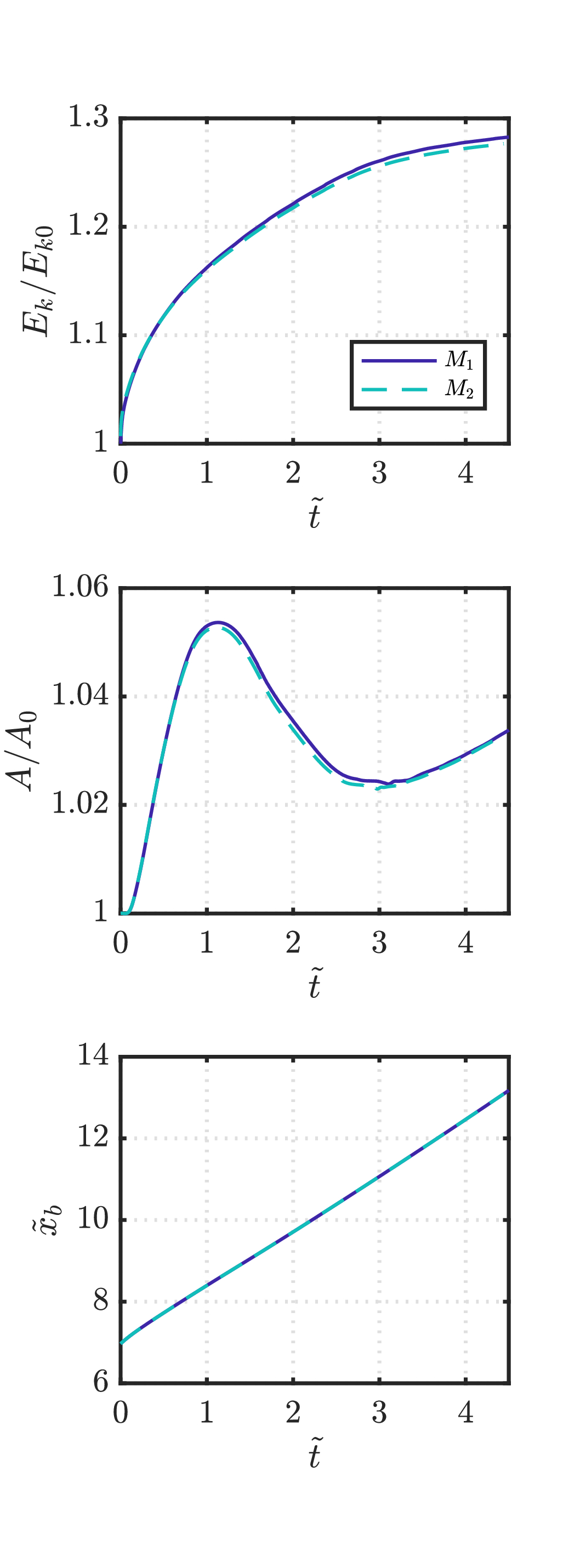}
         \caption{Case 1}
         \label{fig:mesh_case9}
     \end{subfigure}
     \hfill
     \begin{subfigure}[b]{0.328\textwidth}
         \centering
         \includegraphics[width=\textwidth]{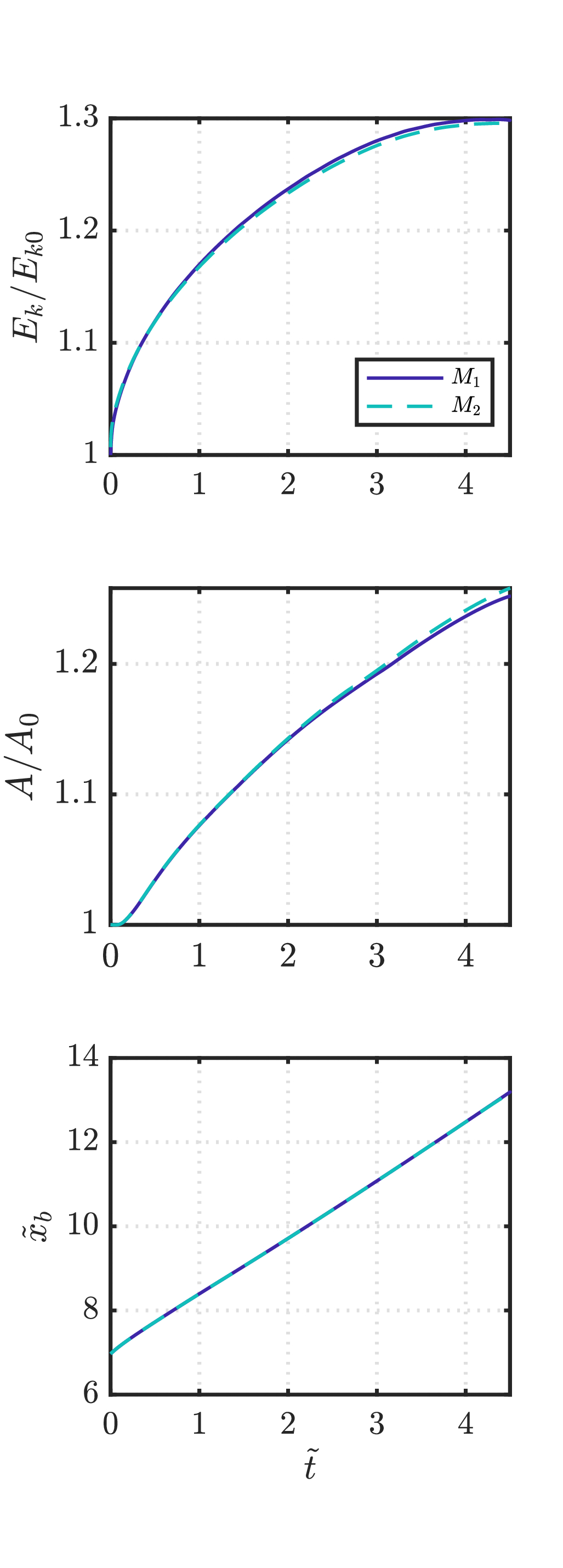}
         \caption{Case 2}
         \label{fig:mesh_case2}
     \end{subfigure}
     \hfill
        \caption{Results of mesh independence study for surfactant-free and surfactant-laden cases}
        \label{fig:mesh}
\end{figure}

For the assessment of our results' (in)dependency on the mesh, we have tested two uniform Cartesian grids with varying resolutions on both surfactant-free and surfactant-laden cases. The operating conditions for all testing cases are those that allow the development of the liquid cavity and drop encapsulation phenomenon, thus evaluating the resolution on small velocity and time scales. Figure \ref{fig:mesh} depicts the temporal evolution of the kinetic energy, $E_{k} = \int_{V}(\rho \textbf{u}^{2})/2 \, dV$, the interfacial area, and the position of the bubble nose, $\tilde{x}_{b}$, the former two being normalised by their values at $\tilde t = 0$. The tested mesh resolutions correspond to $M_{1} = $ $3456$ × $128$ × $128$ and $M_{2} = $ $6912$ × $256$ × $256$, and the cases to surfactant-free for ``case 0'' and surfactant-laden with $Da = 1$, $k = 0.1$ for ``case 1'' and $Da = 0.1$, $k = 1$ for ``case 2''. All other parameters remain unchanged from those specified in §\ref{subsec:sim_setup}. Table \ref{tab:mesh} reports a few additional features of the tested mesh resolutions, including the minimum cell size, $\Delta x_{min}$, the maximum relative variation of the gas phase volume, $\Delta V_{g} = (V_{g} - V_{g0})/V_{g0}$, and that of the liquid phase, $\Delta V_{l} = (V_{l} - V_{l0})/V_{l0}$. The almost complete overlapping of the plots displayed in figure \ref{fig:mesh} for the grids considered and the three sets of operating conditions suggests that increasing the resolution by twofold on each direction does not lead to the development of significantly different dynamics. Unless stated otherwise, all simulations have been run with $M_{1}$ given that it provides enough resolution to ensure mesh-independent results, while also requiring substantially less computational resources, as seen in table \ref{tab:mesh}. This table also indicates that $M_{1}$ allows us to achieve volume conservation for both phases with a deviation of $8\times 10^{-3}$ $\%$ or lower from the onset of the simulations.

\begin{table}
\centering
\caption{Additional features of the mesh resolutions tested}
\label{tab:mesh}
\begin{tabular}{cccccc}
Case name &
  Mesh &
  \begin{tabular}[c]{@{}c@{}}Computing time/CPU\\ to reach $10$ $ms$ (h)\end{tabular} &
  $\Delta x_{min}$ ($\mu m$) &
  $\max(\Delta V_{g})$ (\%) &
  $\max(\Delta V_{l})$ (\%) \\
\multirow{2}{*}{Case 0} & M1 & $\sim$0.60   & 7.81 & $7.70\times10^{-3}$  & $1.04\times10^{-3}$ \\
                        & M2 & $\sim$8.46  & 3.91 & $-6.13\times10^{-3}$ & $8.27\times10^{-4}$ \\
\multirow{2}{*}{Case 1} & M1 & $\sim$0.78  & 7.81 & $7.65\times10^{-3}$  & $3.02\times10^{-4}$ \\
                        & M2 & $\sim$12.71 & 3.91 & $3.84\times10^{-3}$  & $5.19\times10^{-4}$ \\
\multirow{2}{*}{Case 2} & M1 & $\sim$2.21  & 7.81 & $8.00\times10^{-3}$  & $2.61\times10^{-4}$ \\
                        & M2 & $\sim$7.34  & 3.91 & $7.28\times10^{-3}$  & $9.83\times10^{-4}$
\end{tabular}
\end{table}


\bibliographystyle{jfm}
\bibliography{references}
\end{document}